\documentclass{article}

\usepackage{graphicx} 
\usepackage[utf8]{inputenc}
\usepackage[english]{babel}
\usepackage{amsmath}
\usepackage{amsfonts}
\usepackage{amssymb}
\usepackage{graphicx}
\usepackage{a4wide}
\usepackage{xspace}

\graphicspath{{figures/}}

\usepackage{float}
\usepackage{tikz}
\usetikzlibrary{arrows.meta,positioning,shapes.geometric}
\usepackage{physics}
\usepackage{subcaption} 
\usepackage{xcolor}     
\usepackage{booktabs}   
\usepackage{tabu}       
\usepackage{svg}
\usepackage{multirow}
\usepackage[
    backend=biber, style=phys
]{biblatex}
\usepackage{hyperref}
\hypersetup{
    colorlinks,
    linkcolor={red!50!black},
    citecolor={green!50!black},
    urlcolor={blue!80!black}
}
\usepackage[toc,page]{appendix}

\newcolumntype{L}{>{$}l<{$}}
\newcolumntype{C}{>{$}c<{$}}
\newcolumntype{R}{>{$}r<{$}}

\newcommand{\dfthalf}{DFT-$\frac{1}{2}$ }

\newcommand{\supercell}[1]{$#1\times #1 \times #1$}

\title{An accurate DFT-1/2 approach for shallow defect states:\\ Efficient calculation of donor binding energies in silicon}

\author{Joshua Claes$^1$, Bart Partoens$^1$, Dirk Lamoen$^2$, Marcelo Marques$^3$ \& Lara K. Teles$^3$\\
\small $^1$COMMIT, Physics Department, University of Antwerp. Groenenborgerlaan 171, 2020 Belgium\\
\small $^2$EMAT, Physics Department, University of Antwerp. Groenenborgerlaan 171, 2020 Belgium\\
\small $^3$GMSN, Physics Department, Aeronautics Institute of Technology, Sao José dos Campos SP12228-900, Brazil
}

\date{}

\begin{document}

\maketitle

\begin{abstract}

Accurate prediction of shallow-donor electron binding energies is critical for device modeling, dopant activation, and donor-based quantum technologies. Traditional beyond-DFT approaches (e.g., hybrid functionals, GW) are prohibitively expensive for the large supercells needed to capture the extended, hydrogenic wavefunctions, while semi-local DFT underestimates band gaps and suffers from delocalization errors. We present a simple, practical protocol for shallow donors based on the DFT-1/2 approximate quasiparticle correction that maintains the computational cost of standard DFT and enables supercells up to thousands of atoms. This approach provides a straightforward and reproducible workflow that delivers reliable donor binding energies with minimal computational overhead. Applied to group-V donors in Si, Si:X (X$=$ P, As, Sb, Bi), the method yields binding energies in close agreement with experiment. We found that, for Si:Bi, it is essential to include spin-orbit coupling  to achieve near-experimental values with a difference of only $\sim$ 4 meV.  For arsenic, the method yields excellent agreement with experiment, with a difference of only ~0.3 meV. For antimony, the results match experiment to within ~5 meV, and for phosphorus, the deviation is within ~8 meV. Beyond its high accuracy, DFT-1/2 offers a significant practical advantage, providing a straightforward, reproducible, and transferable workflow that is less demanding than hybrid functional approaches while remaining fully generalizable to other shallow impurities in semiconductors.

\end{abstract}

\section{Introduction}

Semiconductors are the foundation of modern electronics, but their technological versatility truly emerges with the deliberate introduction of defects \cite{Aboy2014}. In particular, shallow defects, such as donors and acceptors, play a pivotal role, being crucial for conductivity control in electronic devices, setting carrier activation thresholds and, increasingly, defining energy scales relevant to donor-based quantum technologies \cite{Ma2022}.  A classic example is phosphorus (P) in silicon (Si), in which four of its valence electrons form covalent bonds with neighboring Si atoms, while the fifth occupies the conduction band minimum (CBM), forming a hydrogen-like donor state weakly bound to the P atom due to screening \cite{YuCardona2010fundamentals}. However, providing a reliable theoretical description and accurate prediction of their binding energies, which dictate their activation and functionality, remains a significant challenge. 

Traditional models like Effective Mass Theory (EMT) offer useful qualitative insights \cite{ridley2013quantum} but often lack the quantitative accuracy required for advanced applications, and usually rely on empirical parameters \cite{Zhang2013, Swift2020}.
Furthermore, in particular for silicon, an additional complication arises from the six degenerate valleys at its CBM. These valley states form the hydrogen-like orbitals of the donor system, and due to the $T_d$ symmetry of the doped crystal, this sixfold degeneracy is lifted into $A_1$, $T_2$, and $E$ irreducible representations, which become non-degenerate due to inter-valley coupling \cite{ridley2013quantum, kohn_shallow_donors}. This intricate electronic structure further complicates the accurate theoretical description, particularly for quantitative predictions beyond basic qualitative insights from EMT.

While Density Functional Theory (DFT) is an \textit{ab initio} method capable of inherently describing the chemical environment around a defect, it traditionally struggles with shallow defects \cite{YuCardona2010fundamentals, kohn_shallow_donors,Zhang2013}. This is primarily due to the extended nature of their wavefunctions, which necessitates supercells containing tens of thousands of atoms for an accurate representation \cite{Wang2009potpatch}. Compounding this, standard DFT, particularly with semi-local functionals, suffers from well-known delocalization errors \cite{Sanchez2008_loc_deloc_error, Bryenton2023_deloc_err} and band gap underestimation, leading to unreliable descriptions of the defect wavefunction in the vicinity of the donor. These fundamental shortcomings result in a significant underestimation of binding energies, rendering direct DFT calculations inadequate for quantitative predictions \cite{Zhang2013}. 

 To overcome these deficiencies, beyond-DFT methods, such as hybrid functionals or the GW method, are required, as they offer a more accurate description of electron localization and the band gap. However, these methods impose a significantly high computational cost, especially for large systems. In the last decades, some initiatives were taken to address this issue. 
 For instance, Zhang et al. \cite{Zhang2013} enhanced results from the potential patching method \cite{Wang2009potpatch} by incorporating a 64-atom GW calculation. Meanwhile, Peelaers et al. \cite{Peelaers_2017} simulated shallow states in Si nanowires using hybrid functionals and GW. This was only feasible due to the increased localization of the hydrogenic wavefunction caused by the reduced dimensionality.

In 2020, Swift et al. \cite{Swift2020} tackle this problem by using a tandem approach. with hybrid functional. Combine hybrid functional (HSE) calculations in moderate-sized supercells (up to 1,000 atoms) with standard DFT (PBE) calculations in larger supercells. This technique, which involves extrapolating HSE results to the dilute limit, enabled the accurate estimation of donor binding energies for arsenic and bismuth in silicon, achieving close agreement with experimental values. However, this approach required different computational approaches (e.g., hybrid and standard functionals) or pseudopotential treatments for distinct levels of theory, resulting in distinct wavefunctions between the supercell sizes. Consequently, special care must be taken to consistently extract and combine derived quantities, such as binding energies.

Given the trade-offs between accuracy and computational cost inherent in existing methods, the pursuit of alternative ab initio approaches is essential. The development of a methodology that is both computationally efficient and accurate enough to model shallow defects without resorting to complex, system-dependent combinations of functionals and pseudopotentials is highly welcome. 

 A particularly promising candidate is the \dfthalf method developed by Ferreira, Marques, and Teles  for approximate self-energy corrections within the framework of conventional Kohn-Sham DFT \cite{Ferreira2008,Ferreira2011}. For a detailed overview of the foundational method, we refer to the original papers \cite{Ferreira2008, Ferreira2011}. This approach preserves the favorable computational scaling of standard DFT while correcting its inherent band gap and electron delocalization errors. The \dfthalf method mimics the effect of a quasiparticle correction by subtracting a so-called atomic self-energy potential to the pseudopotential of the DFT calculation. 
While this approach has already proven successful for deep-level defects \cite{dfthalf_NVmin, ChargeTransMn, Efrom_Si_dft12, decoupledDFThalf}, its application to the more delocalized and sensitive shallow defects remains an open question.

In this work, we systematically develop and validate a \dfthalf based methodology specifically tailored for shallow donor states. By incorporating the correction directly into a single, unified DFT calculation, our approach eliminates the need to combine results from different functionals. We demonstrate our method on group V donors in silicon (P, As, Sb, Bi), comparing all results with available experimental data. For As and Bi specifically, we also provide a direct comparison with the computationally expensive tandem hybrid functional calculations of Swift et al. \cite{Swift2020}. We show that our methodology predicts binding energies with great accuracy comparable to experimental data at a significantly reduced computational cost, allowing the use of larger supercells.

\section{Methodology}

\subsection{Calculating the binding energy}

The binding energy ($E_b$) is defined as the energy required to excite an electron from a shallow donor level to the conduction band minimum (CBM). Therefore, it should correspond to the energy difference between the shallow donor level and the CBM. While a single DFT calculation on the doped system technically yields both the donor level and CBM energies, the presence of the defect within a finite supercell significantly perturbs the CBM at the energy scale relevant for binding energies. To circumvent this issue, we determine the CBM from a separate calculation performed on a pristine supercell. The binding energy is then calculated using Eq. \eqref{eq:calc_Eb_DFT} \cite{Swift2020}:

\begin{equation}
    E_b = \varepsilon^{CBM}_\Gamma - \varepsilon^{donor}_\Gamma + e\Delta V \label{eq:calc_Eb_DFT}
\end{equation}

\noindent where $\varepsilon^{CBM}_\Gamma$ represents the CBM energy at the $\Gamma$-point from the pristine supercell, $\varepsilon^{donor}_\Gamma$ is the donor level energy from the doped supercell, and $e\Delta V$ is the electrostatic energy shift correction.

The electrostatic energy shift $\Delta V$ accounts for the change in electrostatic potential introduced by the defect. Since the Kohn-Sham eigenvalues from the doped and pristine calculations no longer share the same reference point, this correction aligns both calculations to a common energy reference \cite{Swift2020, Walle2004FirstprinciplesCF}. We determine $\Delta V$ by constructing a histogram of $\delta V(\vec{r}) = V_\text{bulk}(\vec{r}) - V_\text{defect}(\vec{r})$ sampled at various points $\vec{r}$ within the supercell. The value of $\Delta V$ corresponds to the most frequent bin in this histogram, effectively sampling regions that best resemble the bulk environment \cite{H_CdO_Amini}.

To obtain the binding energy of an isolated shallow donor, we extrapolate results from multiple supercell sizes to the infinite system limit. This is achieved by fitting a linear relationship to data points of $(1/N, E_b(N))$, where $N$ is the number of atoms in the supercell. The binding energy for the isolated donor is taken as the intercept at $1/N = 0$ \cite{Swift2020}. The convergence analysis (detailed in Appendix \ref{apx:sc_conv_Eb}) demonstrates that supercell sizes of \supercell{5} and larger are required for precise binding energy predictions. Therefore, all extrapolations include at least the \supercell{5} and \supercell{6} supercells, with the largest supercell size used for each studied system presented in Tab. \ref{tab:supercell_sizes}.

\begin{table}[h!]
\centering
\caption{Maximum supercell sizes used for each system configuration. The notation indicates: Si (bulk correction), element-1/2 (donor correction), \_d (d-valence orbitals included), SOC (spin-orbit coupling included). HSE results from Swift et al. \cite{Swift2020} are included for comparison.}
\label{tab:supercell_sizes}
\begin{tabular}{lcc}
\toprule
Method & System & SC(max) \\
\midrule
DFT & Si P & \supercell{7} \\
\dfthalf& Si-1/4 P-1/2 & \supercell{8} \\
\midrule
DFT& Si As & \supercell{6} \\
\dfthalf & Si-1/4 As & \supercell{6} \\
\dfthalf & Si-1/4 As-1/2 & \supercell{8} \\
\dfthalf & Si-1/4 As\_d-1/2 & \supercell{7} \\
tandem-HSE & Si As & \supercell{6} \\
\midrule
DFT& Si Sb\_d & \supercell{6} \\
\dfthalf & Si-1/2 Sb-1/2 & \supercell{7} \\
\dfthalf & Si-1/4 Sb\_d-1/2 & \supercell{8} \\
\midrule
DFT & Si Bi & \supercell{6} \\
\dfthalf & Si-1/4 Bi & \supercell{7} \\
\dfthalf & Si-1/4 Bi-1/2 & \supercell{8} \\
\dfthalf & Si-1/4 Bi\_d-1/2 & \supercell{8} \\
\dfthalf & Si-1/4 Bi\_d-1/2 (SOC) & \supercell{6} \\
tandem-HSE & Si Bi & \supercell{6} \\
\bottomrule
\end{tabular}
\end{table}

\begin{figure}[h!]
    \centering
    \includegraphics[width=0.5\linewidth]{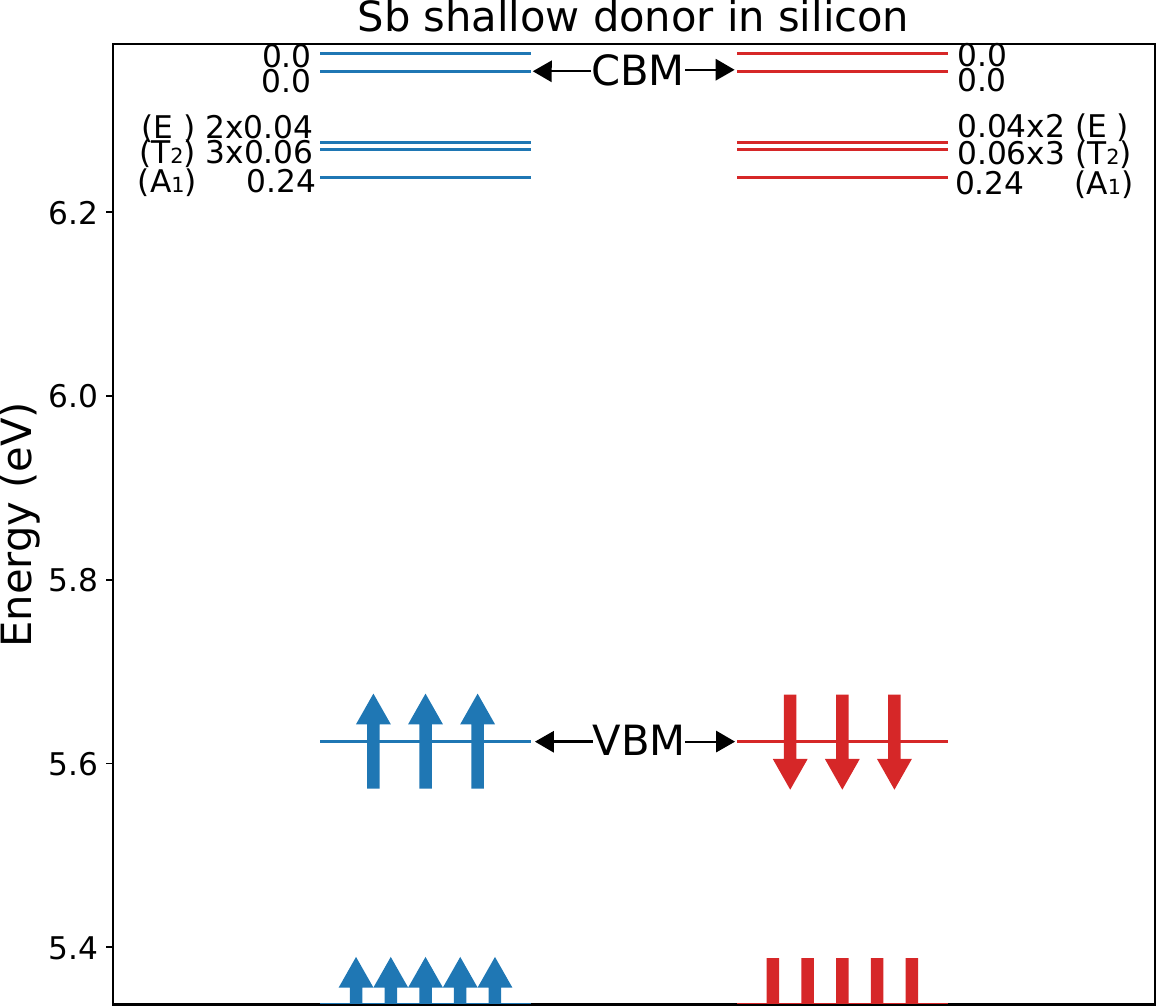}
    \caption{Eigenvalues of a $\Gamma$-point DFT calculation for a shallow Sb donor in Si using high smearing (0.05 eV) for the \supercell{4} supercell. Blue and red bands represent spin-up and spin-down states, respectively. Numbers indicate fractional occupations, while arrows mark fully occupied bands. The figure shows the $A_1$, $T_2$, and $E$ shallow levels along with the VBM and CBM.}
    \label{fig:bs_step1}
\end{figure}

Given the small energy differences between ground state shallow levels, excited shallow levels, and the CBM, special care must be taken regarding the electronic smearing in DFT calculations. We follow the two-step procedure outlined by Swift et al. \cite{Swift2020}. First, an initial DFT calculation is performed using a small smearing value of $5 \times 10^{-4}$ eV. This is followed by a $\Delta$SCF calculation \cite{Gali2009_spin_conserving_ex} to enforce occupation of a single shallow level, ensuring that the ground state shallow level ($A_1$) does not mix with higher-energy shallow levels or conduction bands. Figure \ref{fig:bs_step1} illustrates an example of shallow level mixing for an antimony dopant. We note that with our chosen smearing value in the initial step, the second step rarely changes the outcome for supercell sizes of \supercell{4} or larger.

\subsection{Constructing large supercells}

To determine the binding energy, we did DFT calculations on supercell sizes ranging form the \supercell{2} (64 atoms) to the \supercell{8} (4096 atoms) supercell.
Given the fact that the computational demand of the \dfthalf method is the same of the standard DFT method, the most significant bottleneck is the strucutral relaxation of the large supercell, a step required in any calculation procedure. For large supercells, a single ionic relaxation step can require tens of hours to several days. To overcome this computational challenge while maintaining the accuracy required for shallow defects, whose extended wavefunctions necessitate large supercells, we developed an embedding strategy.  This approach involves placing a pre-relaxed \supercell{5} supercell (1000 atoms) within a pristine silicon environment to construct larger systems (\supercell{6}, \supercell{7}, etc.) without full structural optimization.

\begin{figure}
    \centering
    \begin{subfigure}{0.30\textwidth}
        \includegraphics[height=4.4cm]{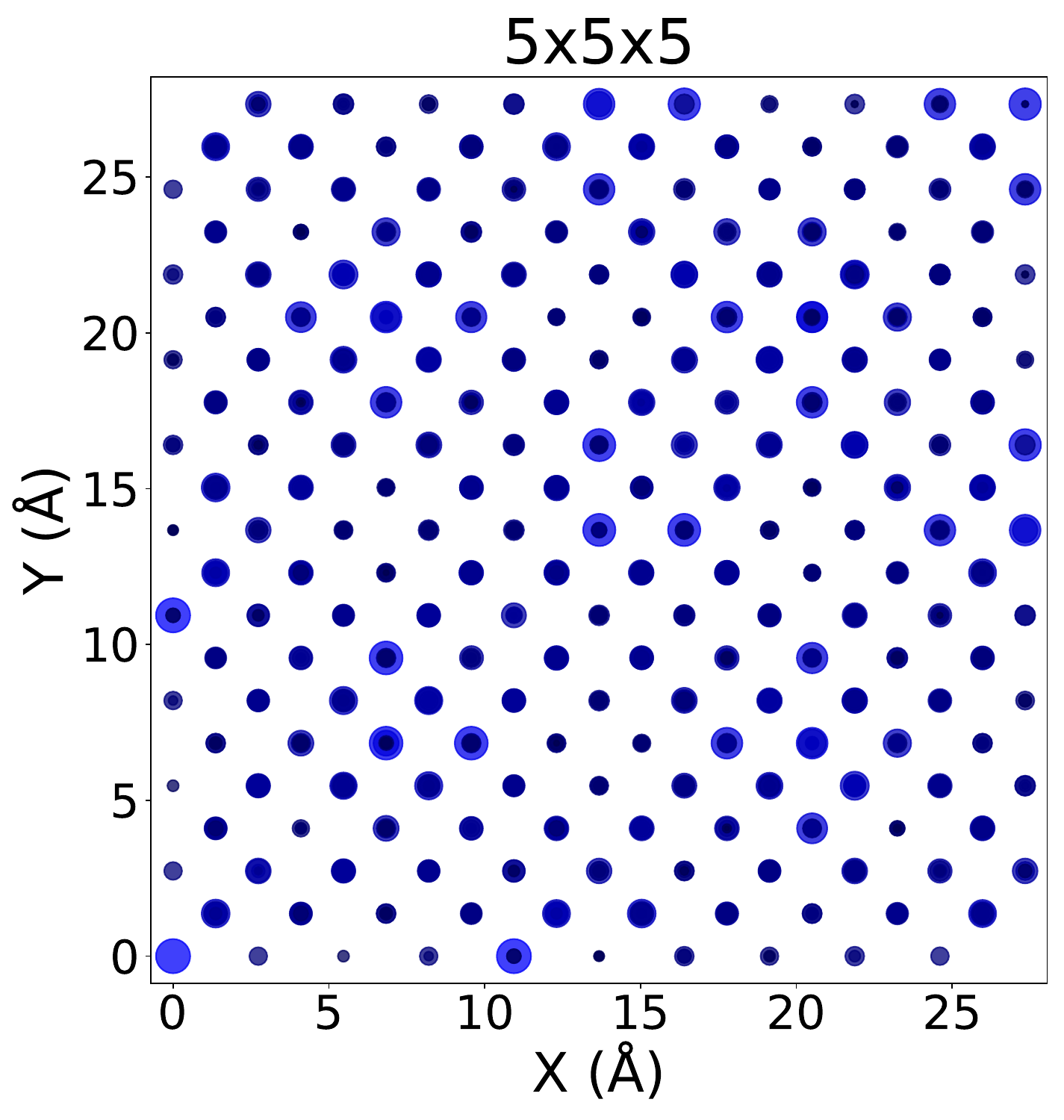}
    \end{subfigure}
        \begin{subfigure}{0.30\textwidth}
        \includegraphics[height=4.4cm]{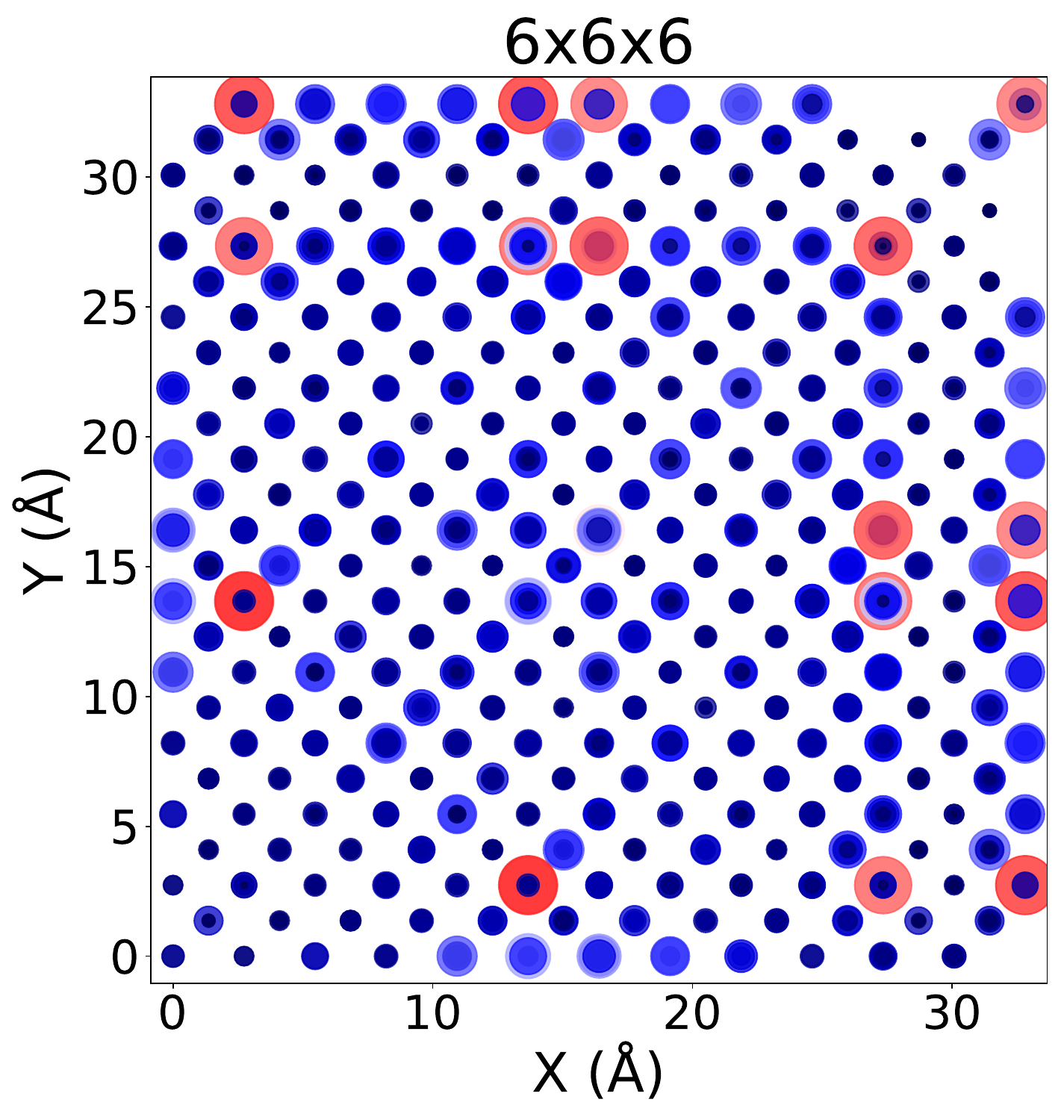}
    \end{subfigure}
        \begin{subfigure}{0.30\textwidth}
        \includegraphics[height=4.4cm]{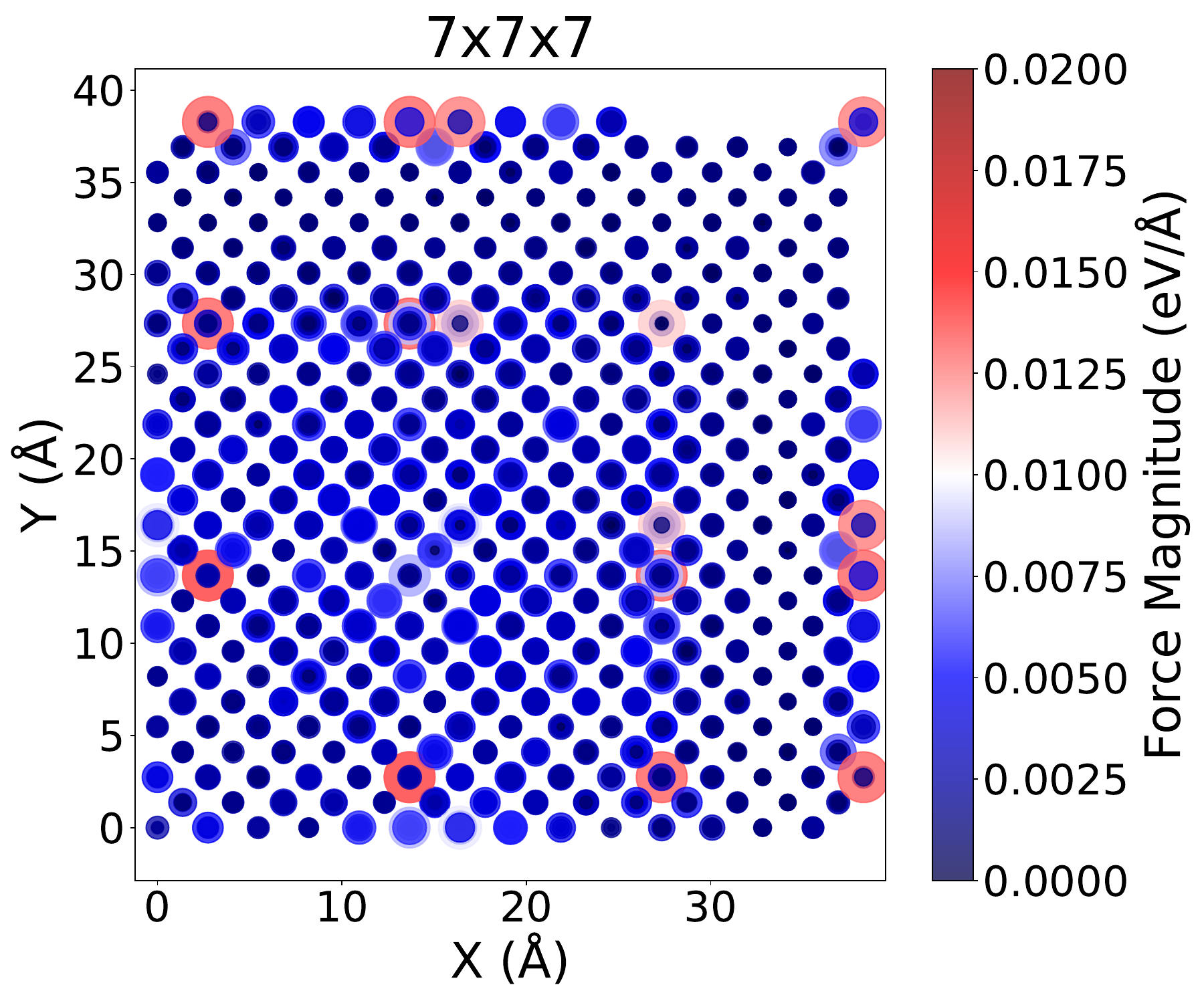}
    \end{subfigure}
    \caption{A projection of the total force on every atom in the $xy$-plane for the \supercell{5} fully relaxed supercell and the larger supercells, \supercell{6} and  \supercell{7}, with embedded \supercell{5} . The forces on each atom are represented by both the color and circle size, with larger circles indicating larger forces.}
    \label{fig:shallow_forces_2d}
\end{figure}

Fig. \ref{fig:shallow_forces_2d} depicts the maximum force acting on each atom within both the original, relaxed \supercell{5} supercell and the embedded \supercell{5} supercells within pristine silicon. We observe that at the interface between the relaxed supercell and the pristine silicon, the forces are indeed larger than those seen in the non-embedded case. However, the largest forces in the supercell is $0.014 \text{eV}/\text{\AA}$  and $0.013 \text{eV}/\text{\AA}$ for the \supercell{6}  (1728 atoms) and \supercell{7} (2744 atoms),  respectively. Although these values are one order of magnitude larger than the original convergence criterion of $10^{-3} \text{eV}/\text{\AA}$, they are comparable to the $10^{-2} \text{ eV}/\text{Å}$ threshold commonly employed in defect studies \cite{Gali_thiering_2018_g4v,ed_scattering_Si,PhysRevLett.123.127401, Ma2022}. Furthermore, previous studies  \cite{effect_rad_def_Si} have indicated that the displacement of silicon atoms due to the creation of a vacancy  becomes negligible when  going from the \supercell{4} to the \supercell{5} supercells.

\subsection{Computational details}\label{sec:comp_details}
The DFT calculations for this work utilized the Perdew-Burke-Ernzerhof (PBE) exchange correlation functional \cite{PBE_functional} and the Projector Augmented Wave (PAW) method \cite{vasppaw}, as implemented in the Vienna Ab initio Simulation Package (VASP) \cite{vaspcode1,vaspcode2,vaspcode3}. Collinear spin polarization was included in all calculations unless otherwise noted. The lattice parameter of bulk silicon was determined using the Birch-Murnaghan equation of state \cite{BMeq}, employing an energy cutoff of $246$ eV and a dense $k$-point grid of \supercell{21} in the Monkhorst-Pack scheme \cite{Monkhorst_Pack_scheme}. This yielded silicon-silicon interatomic distances and a lattice parameter of $2.37$ $\text{\AA}$ and $5.47$ \AA, respectively, which is in good agreement with the experimentally determined lattice parameter of $5.43$ $\text{\AA}$ \cite{semiconductor_properties}. The \dfthalf correction was implemented following the procedure outlined by Ferreira et al.  \cite{Ferreira2008}. We removed 1/4th of an electron from the $p$-orbital of silicon and determined the optimal cutoff parameter by extremizing the band gap. By using a dense \supercell{21} $k$-point grid, we obtained a cutoff parameter of $3.7$ $a_0$ and a \dfthalf band gap of $1.3$ eV. These results align with previous \dfthalf calculations \cite{Ferreira2008, limitsdfthalf}, and the band gap is close to the experimentally measured value of $1.17$ eV \cite{semiconductor_properties}. The Kohn Sham potentials used to calculate the \dfthalf self-energy potentials were generated using a modified version of the \textsc{atom} code \cite{Siesta_package,Ferreira2008}.

For defect calculations, we employed different $k$-point sampling schemes depending on supercell size: $3 \times 3 \times 3$ for \supercell{2} (64 atoms), $2 \times 2 \times 2$ for \supercell{3} (216 atoms), and $\Gamma$-point only for larger supercells. The energy cutoffs for each donor species and their respective pseudopotentials are listed in Table \ref{tab:shallow_encuts}. For structural relaxations, we used energy cutoffs scaled by a factor of 1.3 relative to the values in Table \ref{tab:shallow_encuts} to ensure accurate forces and stress tensors.

\begin{table}[h]
    \centering
    \caption{Energy cutoffs and pseudopotential specifications for each donor species. The \_d suffix indicates that $d$-orbitals are treated as valence electrons, while \_GW indicates the use of GW-optimized pseudopotentials. Valence configurations show the electrons explicitly treated in the calculations.}
    \label{tab:shallow_encuts}
    \begin{tabular}{lccc}
    \toprule
        Donor & POTCAR type & Valence config. & ENCUT (eV)\\ 
    \midrule
        P & standard & $3s^2 3p^3$ & 256\\
        As & standard & $4s^2 4p^3$ & 246\\
        As & \_d & $3d^{10} 4s^2 4p^3$ & 289\\
        Sb & standard & $5s^2 5p^3$ & 246\\
        Sb & \_GW\_d & $4d^{10} 5s^2 5p^3$ & 289 \\
        Bi & standard & $6s^2 6p^3$ & 246 \\
        Bi & \_d & $5d^{10} 6s^2 6p^3$ & 246\\ 
    \bottomrule
    \end{tabular}
\end{table}

\section{Results and Discussion}

\begin{figure}[h!]
    \centering
    \begin{subfigure}{0.49\textwidth}
        \centering
        \includegraphics[width=\linewidth]{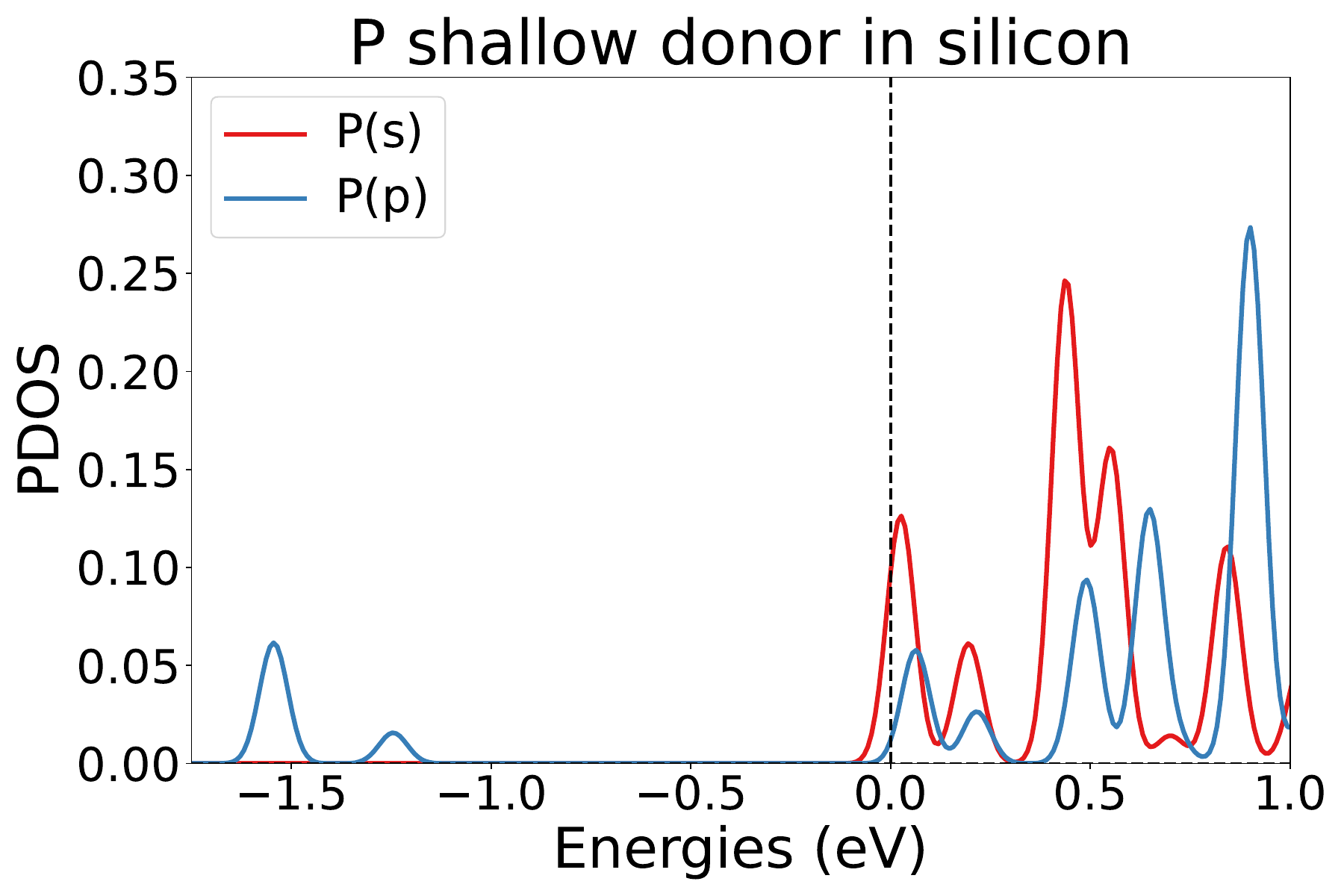}
        \caption{}
        \label{fig:P_pdos}
    \end{subfigure}
    \hfill
    \begin{subfigure}{0.49\textwidth}
        \centering
        \includegraphics[width=\textwidth]{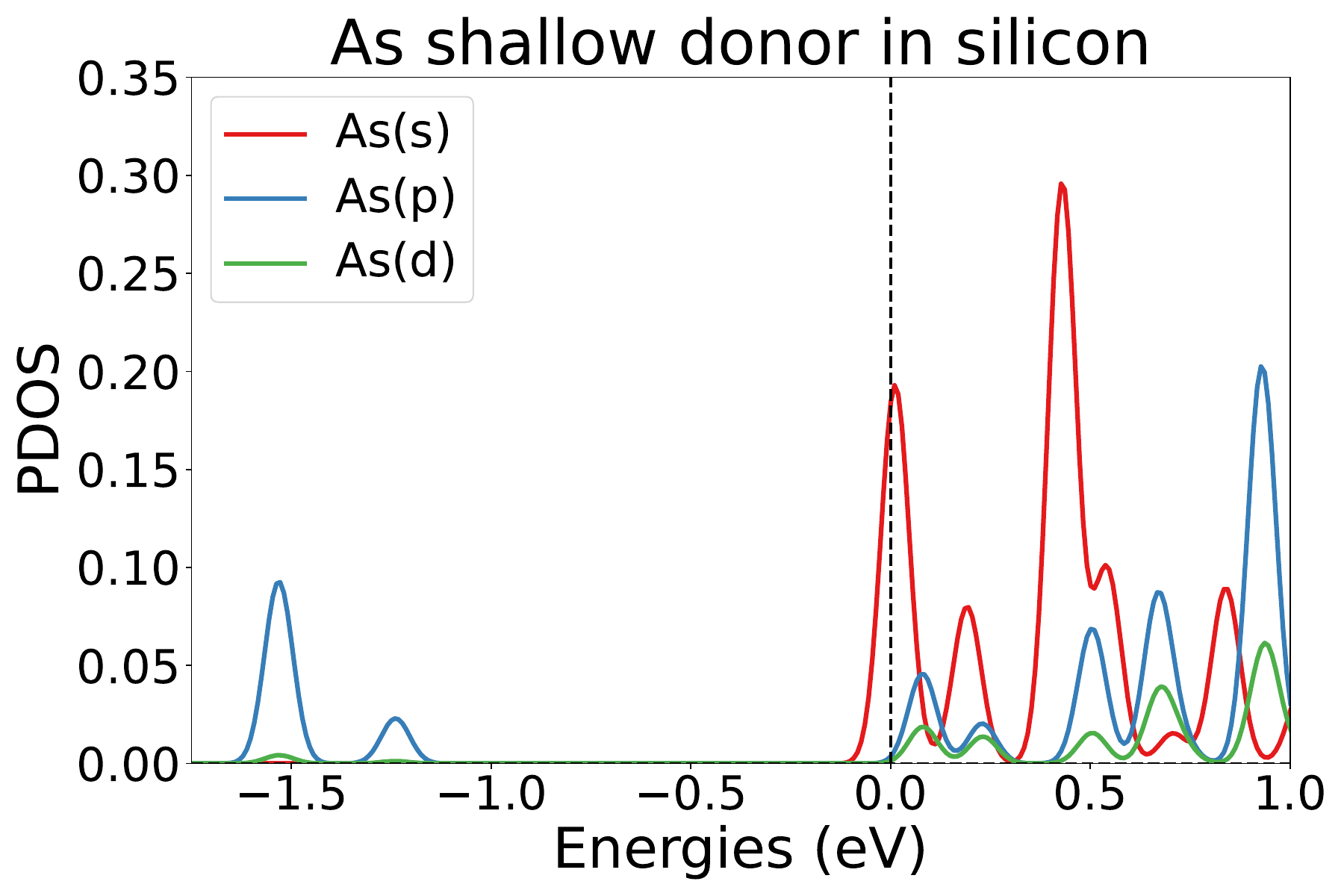}
        \caption{}
        \label{fig:As-pdos}
    \end{subfigure}
    
    \begin{subfigure}{0.49\textwidth}
        \centering
        \includegraphics[width=\linewidth]{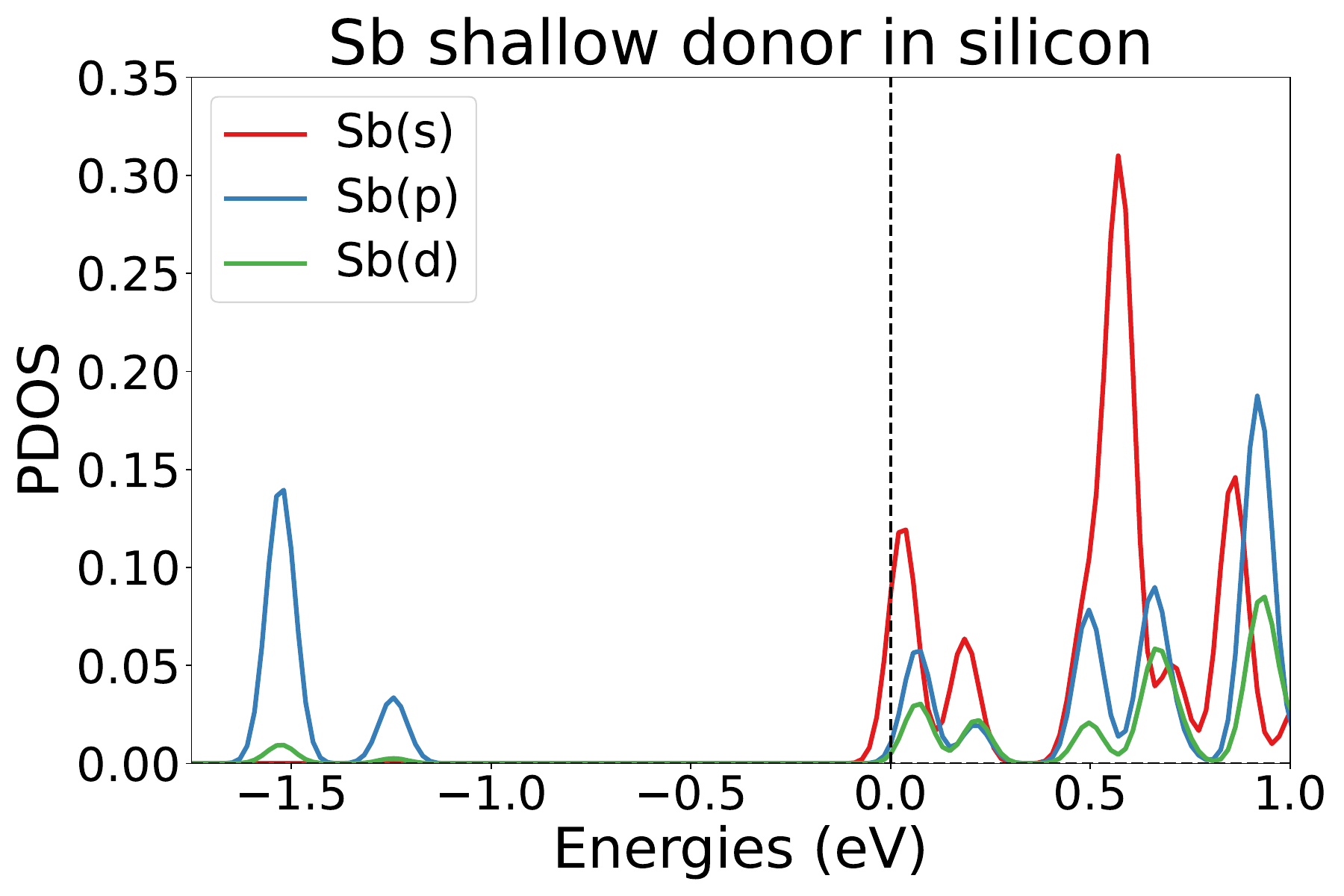}
        \caption{}
        \label{fig:P_pdos}
    \end{subfigure}
    \hfill
    \begin{subfigure}{0.49\textwidth}
        \centering
        \includegraphics[width=\textwidth]{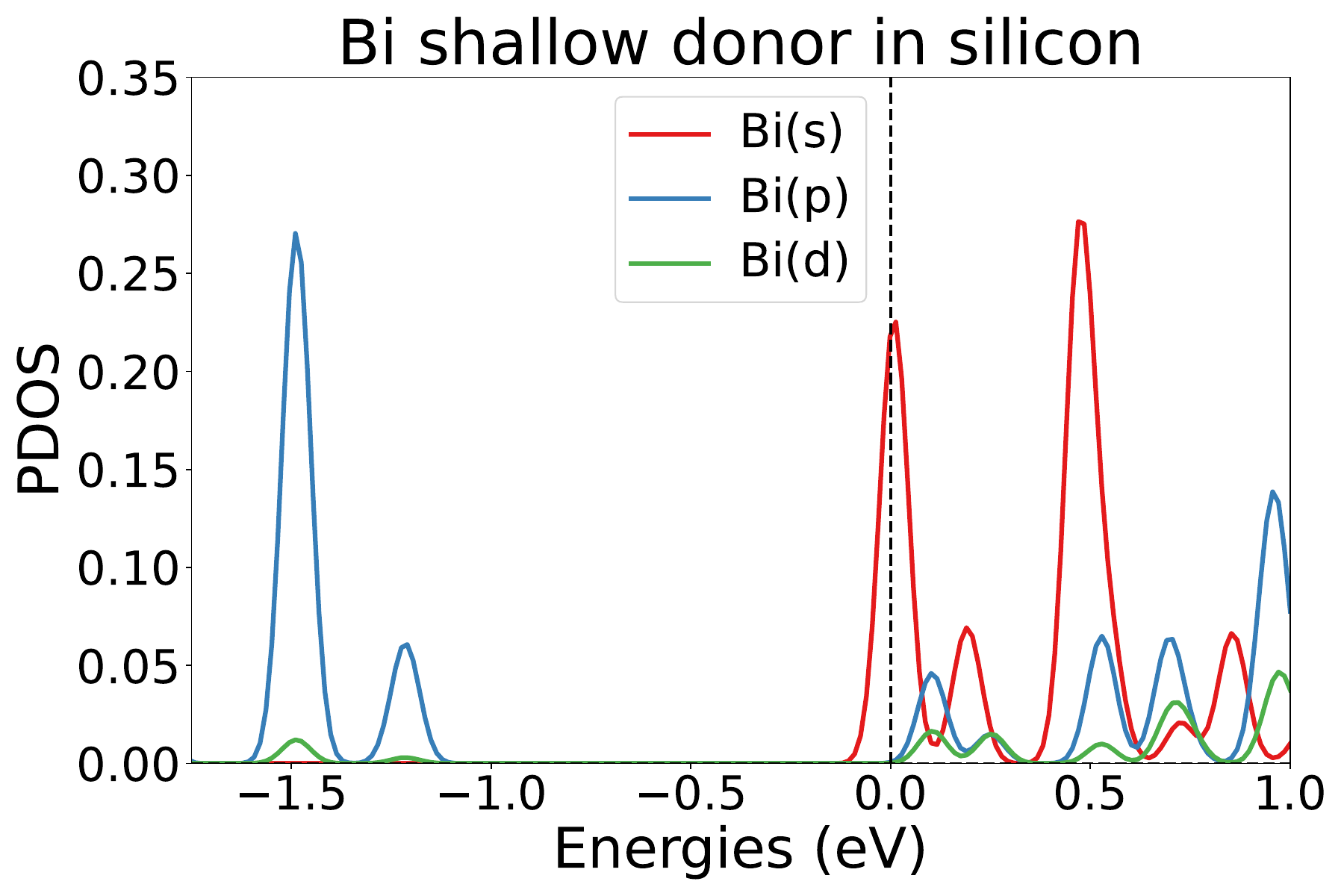}
        \caption{}
        \label{fig:Bi_pdos}
    \end{subfigure}
    \caption{The PDOS for all the group V donors in silicon for a \supercell{4} supercell. For the creation of the PDOS the smearing was increased to 0.05 eV and a \dfthalf correction was applied to the silicon atoms. Due to the shallow donor the fermi level is close to the CBM.
    }
    \label{fig:PDOS}
\end{figure}

\begin{figure}[h!]
    \centering
    \begin{subfigure}{0.49\textwidth}
        \centering
        \includegraphics[width=\textwidth]{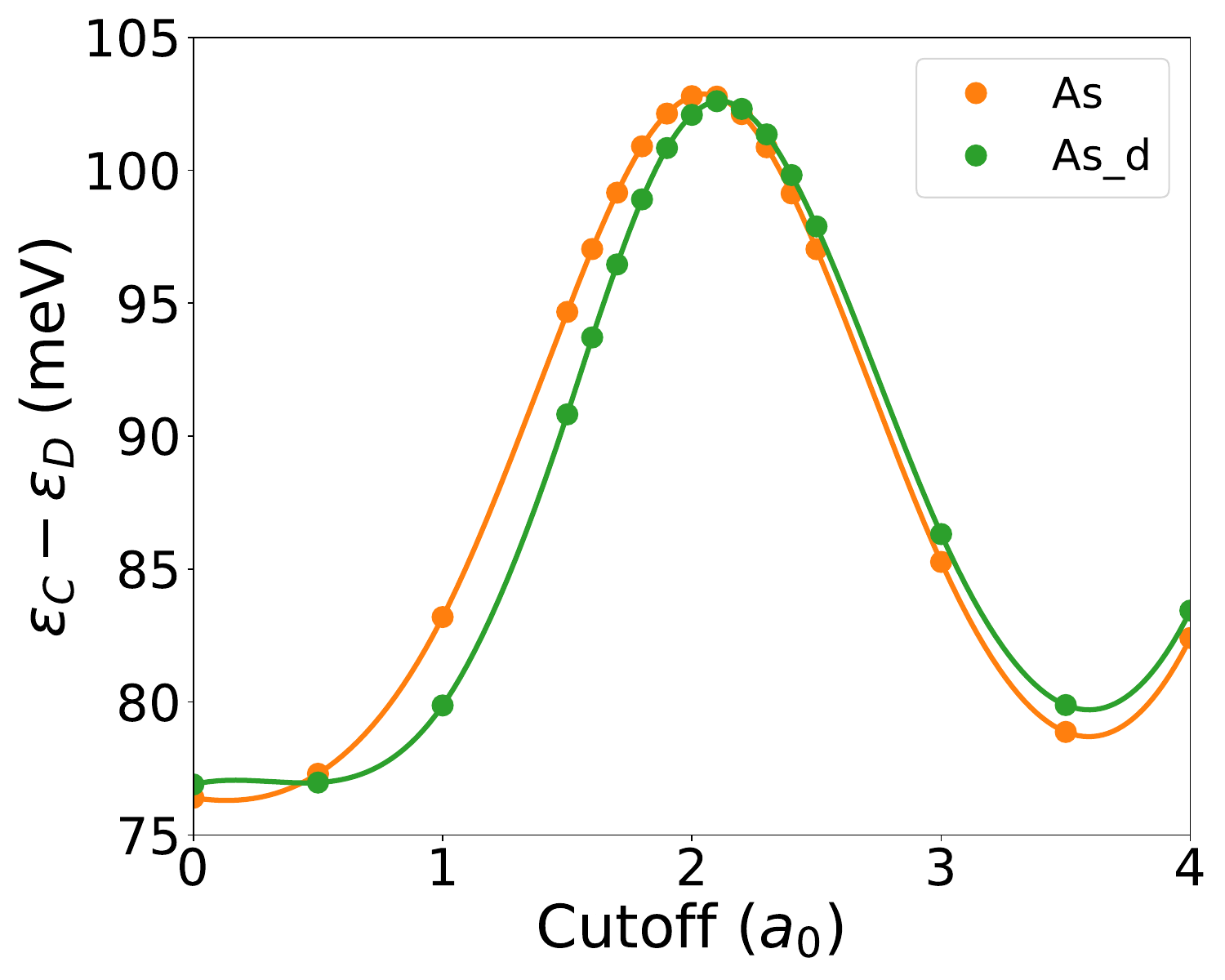}
        \caption{}
        \label{fig:As_cutoff}
    \end{subfigure}
    \hfill
    \begin{subfigure}{0.49\textwidth}
        \centering
        \includegraphics[width=\textwidth]{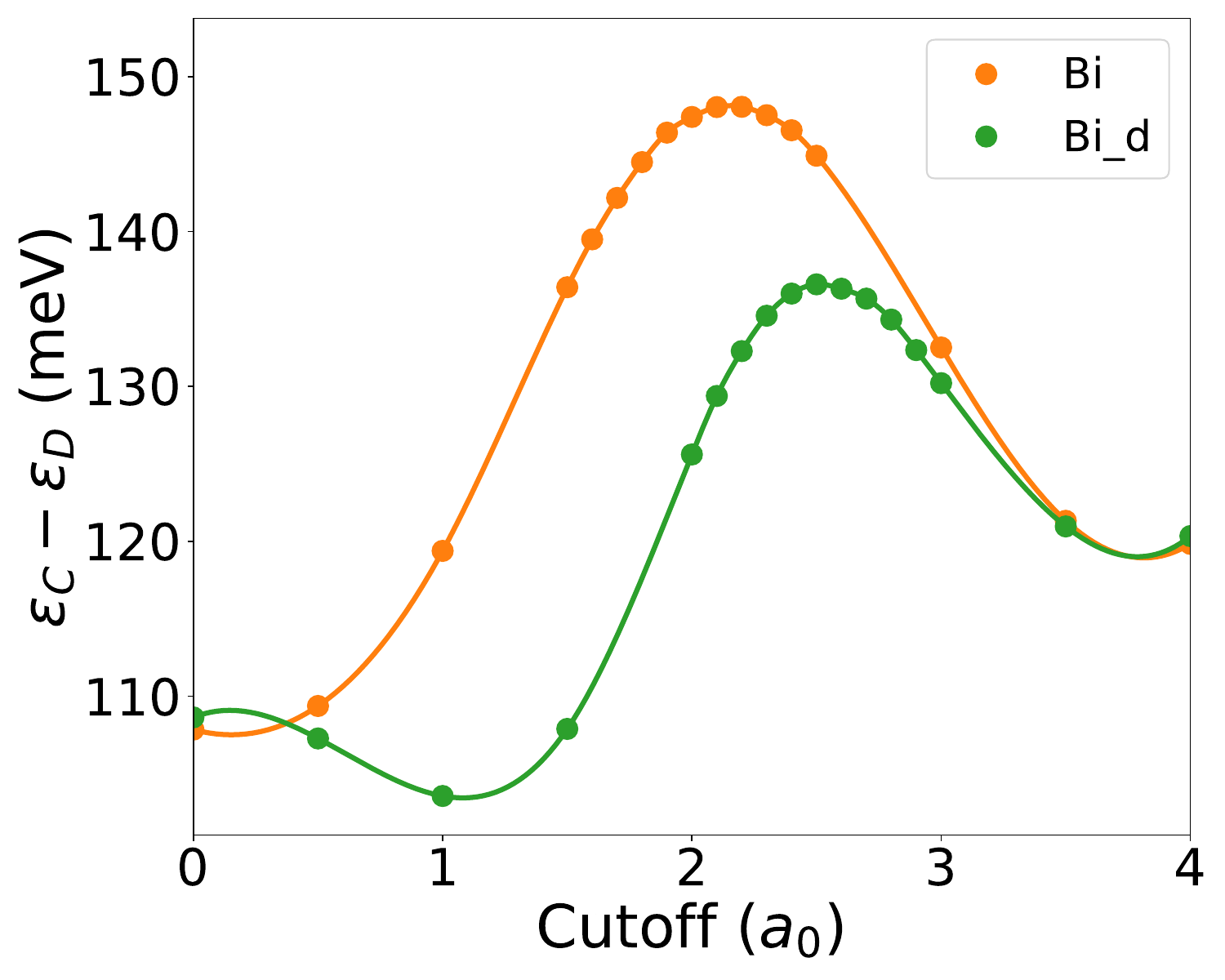}
        \caption{}
        \label{fig:Bi_cutoff}
    \end{subfigure}
    \caption{The \dfthalf cutoff optimization for As and Bi donor in silicon in both case half an electron is removed from the $s$-orbital. The optimized energy separation $\varepsilon_C-\varepsilon_D$, represents the energy difference between  the first empty conduction band and the occupied shallow level  of the same spin. Optimization using a pseudopotential with $d$-valence orbitals is indicated by the labels As\_d and Bi\_d.
    }
    \label{fig:shallow_cutoff_sweeps}
\end{figure}

First, we apply the bulk \dfthalf correction to pristine silicon, resulting in an increased band gap of $1.3$ eV, which is in good agreement with the experimental value of $1.17$ eV, as discussed in Section \ref{sec:comp_details}. This correction is essential as it establishes the proper energy reference for the conduction band minimum, directly impacting the accuracy of binding energy calculations. 
As a second step, we focus on the donor orbital analysis and its correction, in which the bulk \dfthalf correction for silicon has already been incorporated. For this, we analyze the character of the highest occupied level, in this case, the impurity (shallow donor) level. To this end, we compute the projected density of states (PDOS) analysis shown in Figure \ref{fig:PDOS}. It reveals that for all group V donors, the shallow donor state exhibits predominantly $s$-character from the donor atom. This physical insight guides our application of the \dfthalf correction; we remove 1/2 electron from the $s$-orbital of the donor atom, obtain the self-energy potential by optimizing the cutoff parameter by maximizing the energy separation between the first empty conduction band and the occupied shallow level of the same spin. Denoting the shallow donor level by $\varepsilon_D$  and the nearest unoccupied conduction band level of the same spin by $\varepsilon_C$, the optimal cutoff radius ($r_c$) is determined by maximizing the energy separation $\varepsilon_C-\varepsilon_D$. The resulting optimization curves for different cutoff radii are shown, e.g.,  in Fig. \ref{fig:shallow_cutoff_sweeps} for As and Bi within a \supercell{4} supercell. An important practical consideration is the transferability of our optimized \dfthalf potential. As demonstrated in Appendix \ref{Apx:dfthalf_cutoff_conv},  the cutoff parameters determined for \supercell{4} remain optimal for larger supercells, confirming the robustness of the \dfthalf method. This  transferability is crucial for practical applications, as it eliminates the need for re-optimization when scaling to larger systems.


\begin{figure}[h!]
    \centering
    \begin{subfigure}{0.49\textwidth}
        \centering
        \includegraphics[width=\linewidth]{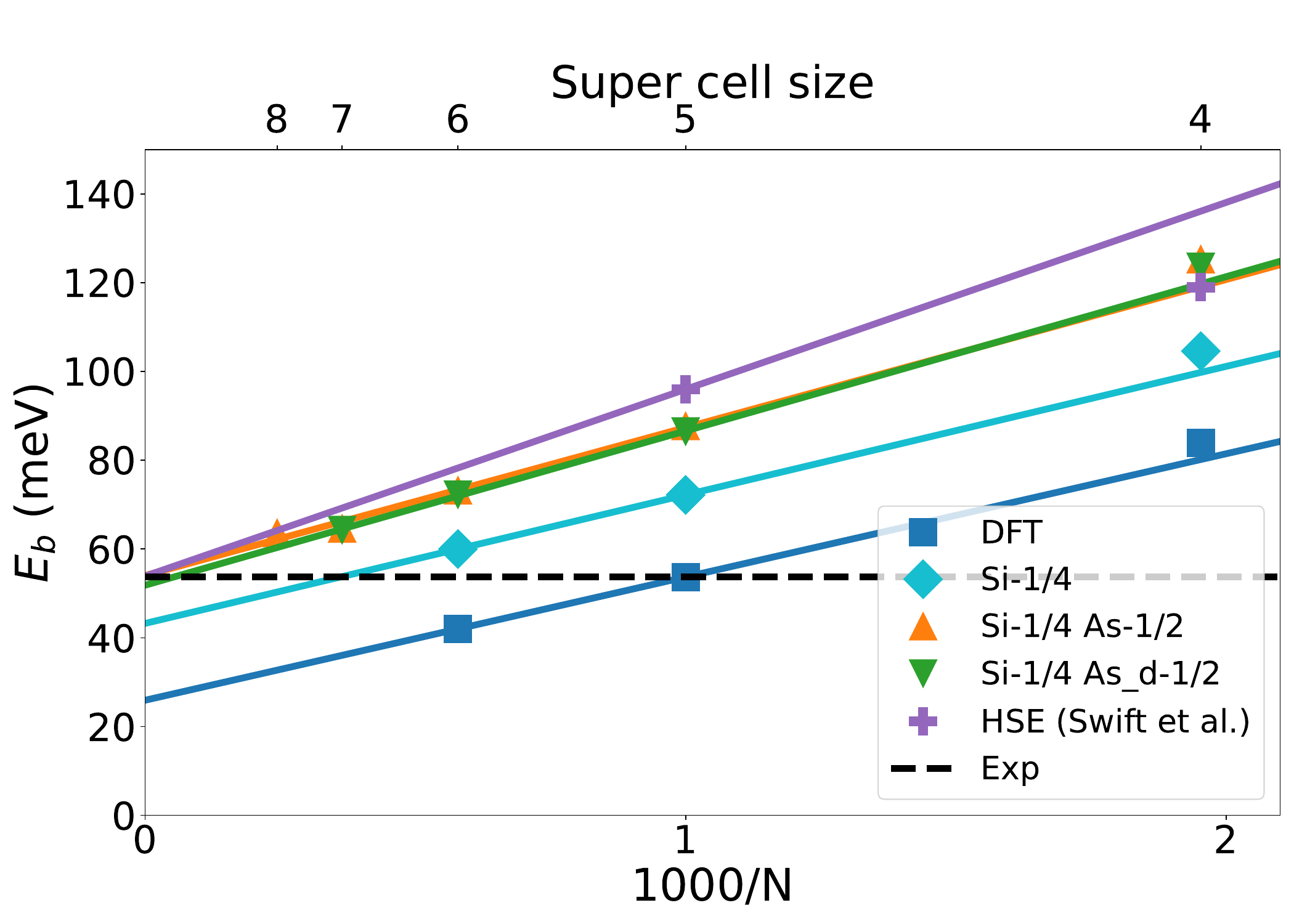}
        \caption{As}
        \label{fig:As_lin_fits}
    \end{subfigure}
    \hfill
    \begin{subfigure}{0.49\textwidth}
        \centering
        \includegraphics[width=\textwidth]{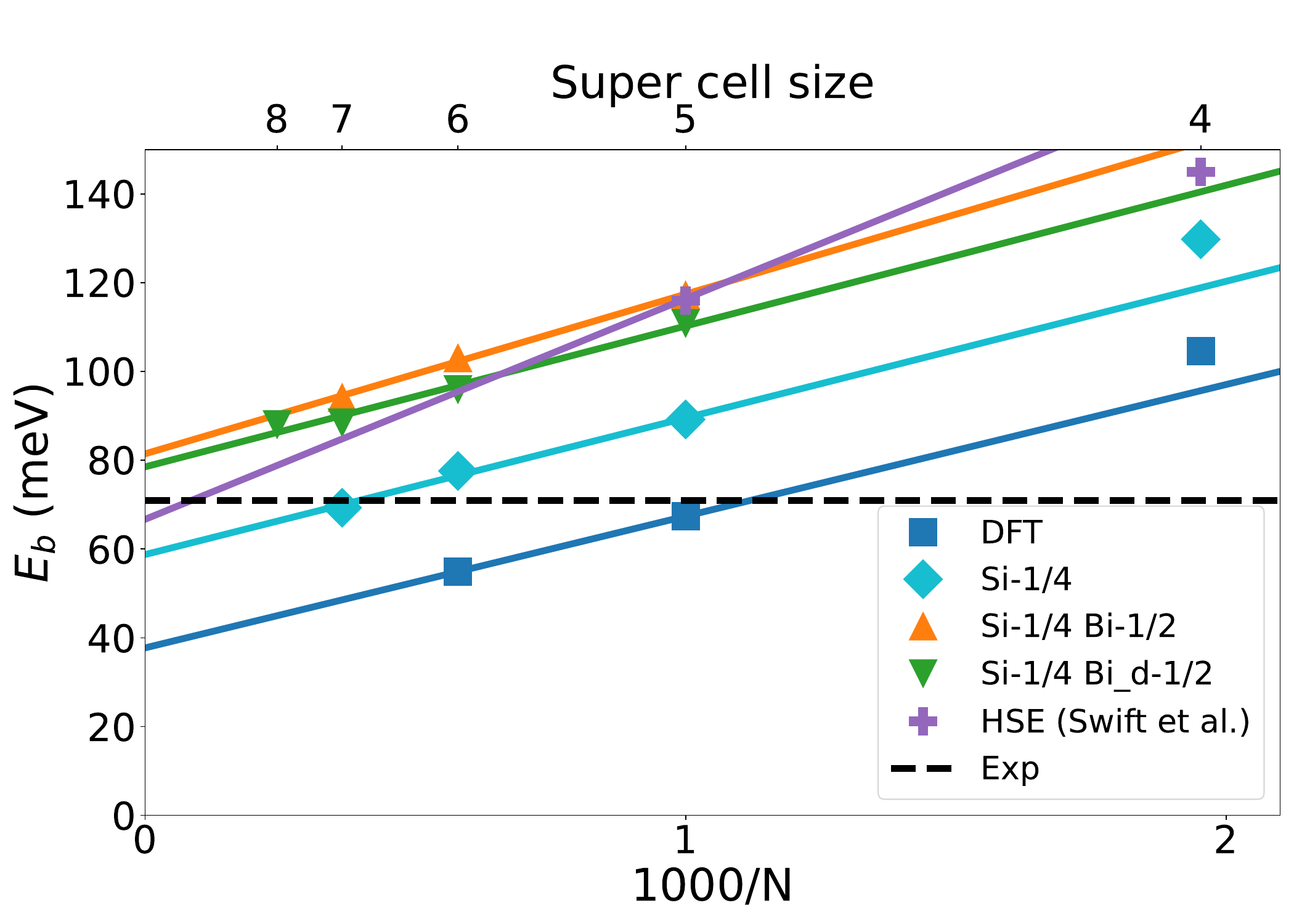}
        \caption{Bi}
        \label{fig:Bi_lin_fits}
    \end{subfigure}
    
    \begin{subfigure}{0.49\textwidth}
        \centering
        \includegraphics[width=\linewidth]{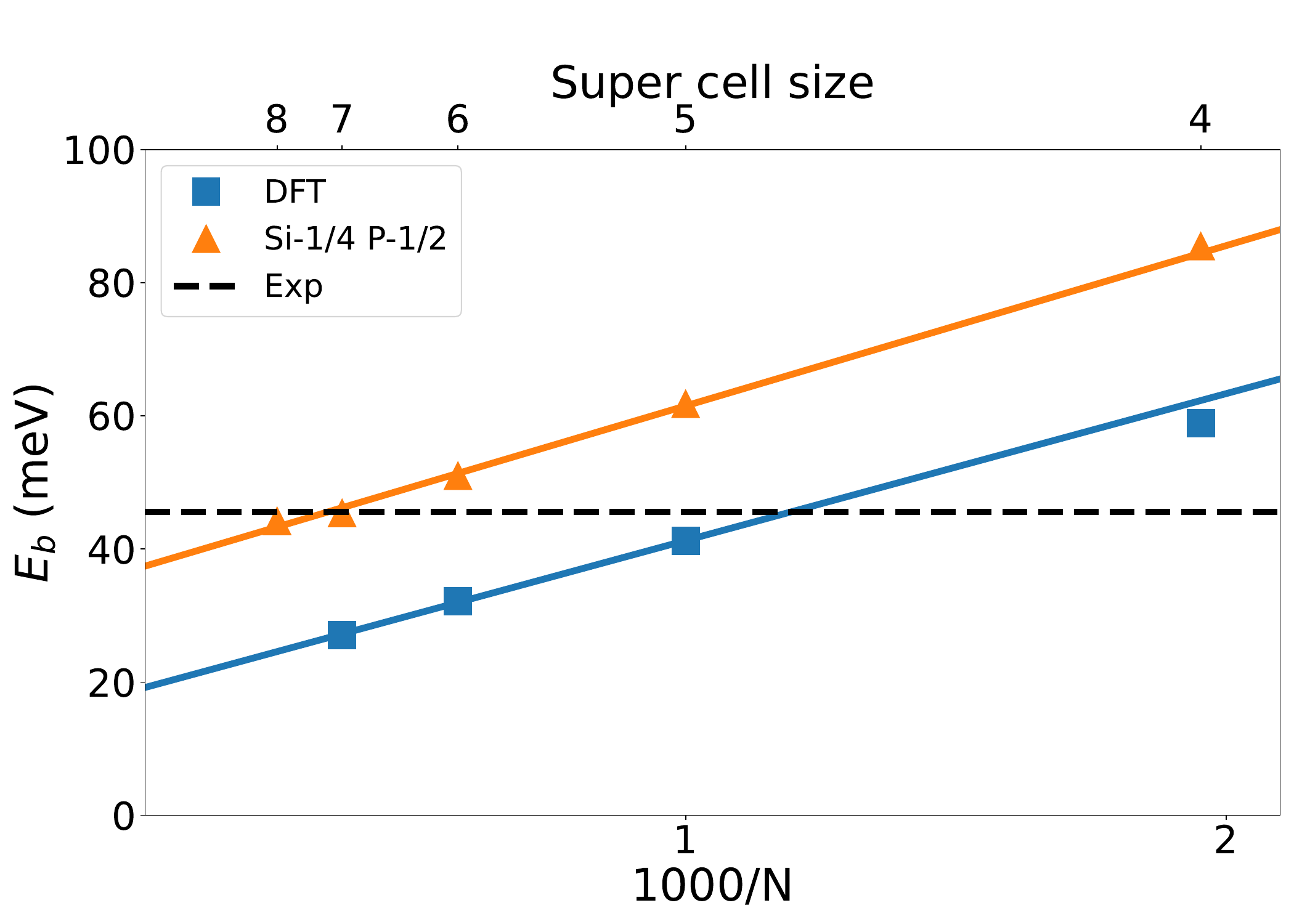}
        \caption{P}
        \label{fig:P_lin_fits}
    \end{subfigure}
    \hfill
    \begin{subfigure}{0.49\textwidth}
        \centering
        \includegraphics[width=\textwidth]{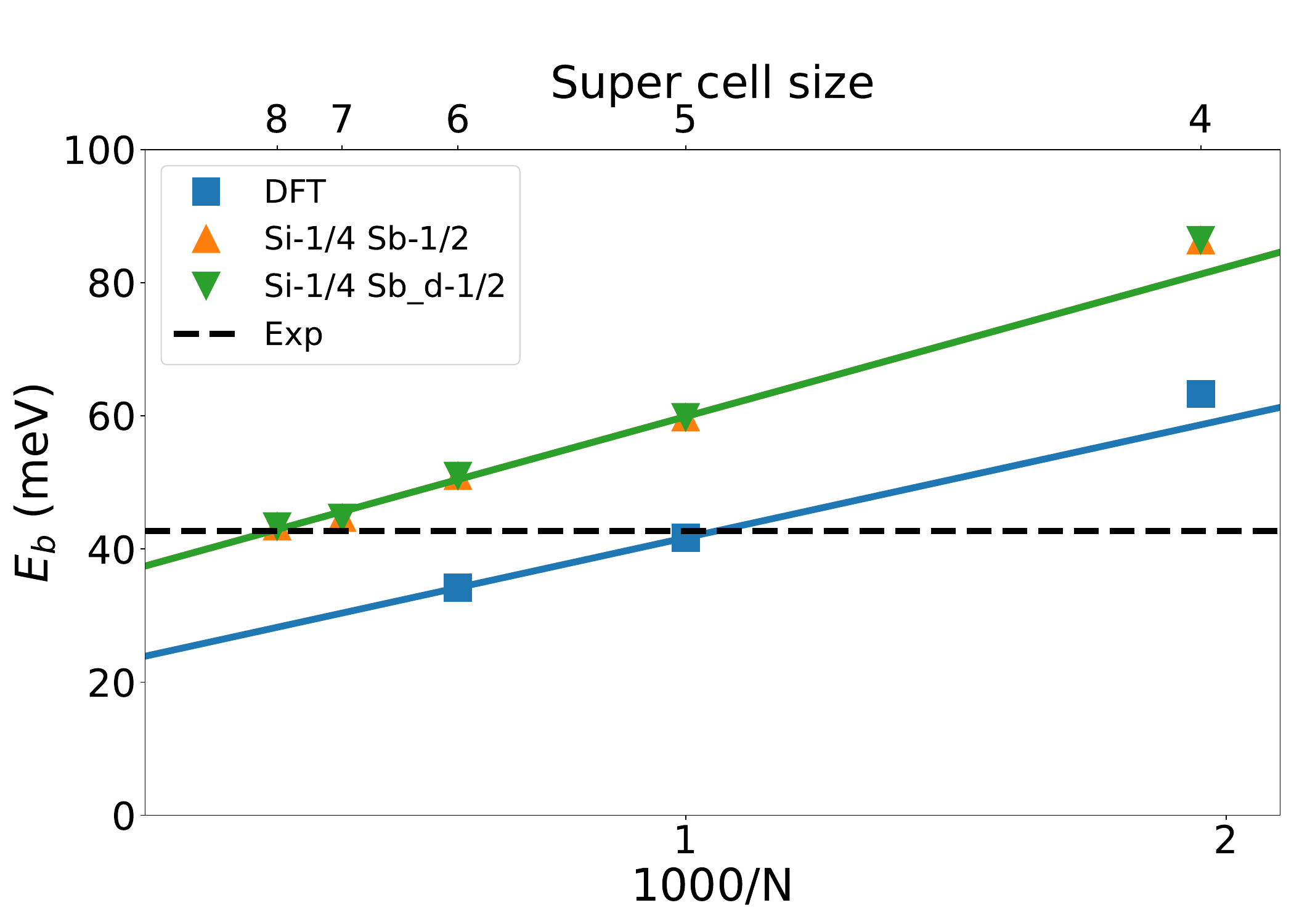}
        \caption{Sb}
        \label{fig:Sb_lin_fits}
    \end{subfigure}
    \caption{The binding energy for group V donors in silicon at various supercell sizes across different levels of the \dfthalf correction along with fit lines is shown. The fits in these figures are made with all binding energies of supercell sizes of \supercell{5} and larger. Binding energies calculated with only the bulk \dfthalf correction are denoted by Si$-\frac{1}{4}$. The plots for arsenic and bismuth incorporate the HSE data and fit from \cite{Swift2020}. The experimental results are obtained from Tab. 1 in Ref. \cite{Saraiva_2015}.}
    \label{fig:shallow_lin_fits}
\end{figure}

In Fig. \ref{fig:As_lin_fits} and \ref{fig:Bi_lin_fits}, the binding energy is calculated for different supercell sizes, along with an extrapolation to the infinite supercell using both standard DFT, hereafter referred to simply as DFT, and \dfthalf, considering the different steps of correction, first with only the bulk silicon correction, denoted as Si$-\frac{1}{4}$, and after incorporating the effect of the donor atom, denoted as Si$-\frac{1}{4}X-\frac{1}{2}$ ($X=$ As, Bi, P, Sb). As expected, with unmodified PBE pseudopotentials, all binding energies are systematically underestimated  with respect to the experimental data (e.g., P: \(19.2\)~meV vs.\ \(45.6\)~meV; As: \(25.9\)~meV vs.\ \(53.8\)~meV; Sb: \(23.9\)~meV vs.\ \(42.7\)~meV; Bi: \(37.7\)~meV vs.\ \(70.9\)~meV), due to the excessive delocalization of the wave functions inherent to the PBE functional. 
The \dfthalf correction on the host silicon potential increases the band gap, which  results in an increase in the level separation between the shallow level and the CBM. This in turn results in a substantial increase of the binding energy, bringing the prediction closer to the experimental results,  (e.g., As: \(43.24\)~meV; Bi: \(58.76\)~meV), while correcting also the impurities brings the results into near-quantitative agreement for lighter donors and  improves the heavier ones. For arsenic, the combined scheme Si-1/4 As-1/2 yields \(E_b=54.04\)~meV, in excellent agreement with the experiment (\(53.77\)~meV) and indistinguishable from the tandem-HSE benchmark (\(53.9\)~meV) \cite{Swift2020}. The prediction for phosphorus shows a marked improvement over standard DFT, with a $\sim$8 meV deviation from experiment.  Previous work employing the tandem-HSE approach for phosphorus \cite{Ma2022} reports a prediction that closely matches the experimental value. However, as discussed in Appendix~\ref{apx:HSE_P_literature}, we argue that this apparent agreement likely results from an overestimation of the electrostatic potential alignment term, $e \Delta V$. If this contribution were corrected, the HSE result would likely align with our \dfthalf prediction. For the heavy donors, additional physics becomes important. 


For Sb and Bi,  we systematically investigate the effects of including $d$-orbitals as valence electrons. It reveals a clear trend, treating $d$-orbitals as valence electrons consistently improves accuracy by 2-3 meV. When including $d$-valence orbitals for arsenic, the lighter element, the effect is negligible in either the cutoff parameter or the final binding energy. This behavior reflects the increasing spatial extent of $d$-orbitals with atomic number, for Sb and Bi, the $d$-orbitals are sufficiently extended to participate in bonding interactions with the silicon host. For bismuth, one can observe an improvement; the binding energy changes from 81.46 meV (standard pseudopotential) to 78.52 meV ($d$-valence), bringing the result 3 meV closer to the experiment. For antimony, an improvement of 2 meV is observed, with a final result underestimated by only $\sim 5$~meV. The cutoff optimization curves in Figure \ref{fig:Bi_cutoff} clearly show how $d$-orbital inclusion affects the electronic structure, the optimal cutoff parameter shifts, and the maximum energy separation $\varepsilon_C-\varepsilon_D$ changes, reflecting the altered electronic environment around the donor atom. Fig. \ref{fig:shallow_lin_fits} and Tab. \ref{tab:binding_energies} illustrate the new binding energies for all materials comparing the steps of each correction for comparison.

\begin{figure}[h!]
    \centering
    \begin{subfigure}{0.40\textwidth}
        \centering
        \includegraphics[width=\linewidth]{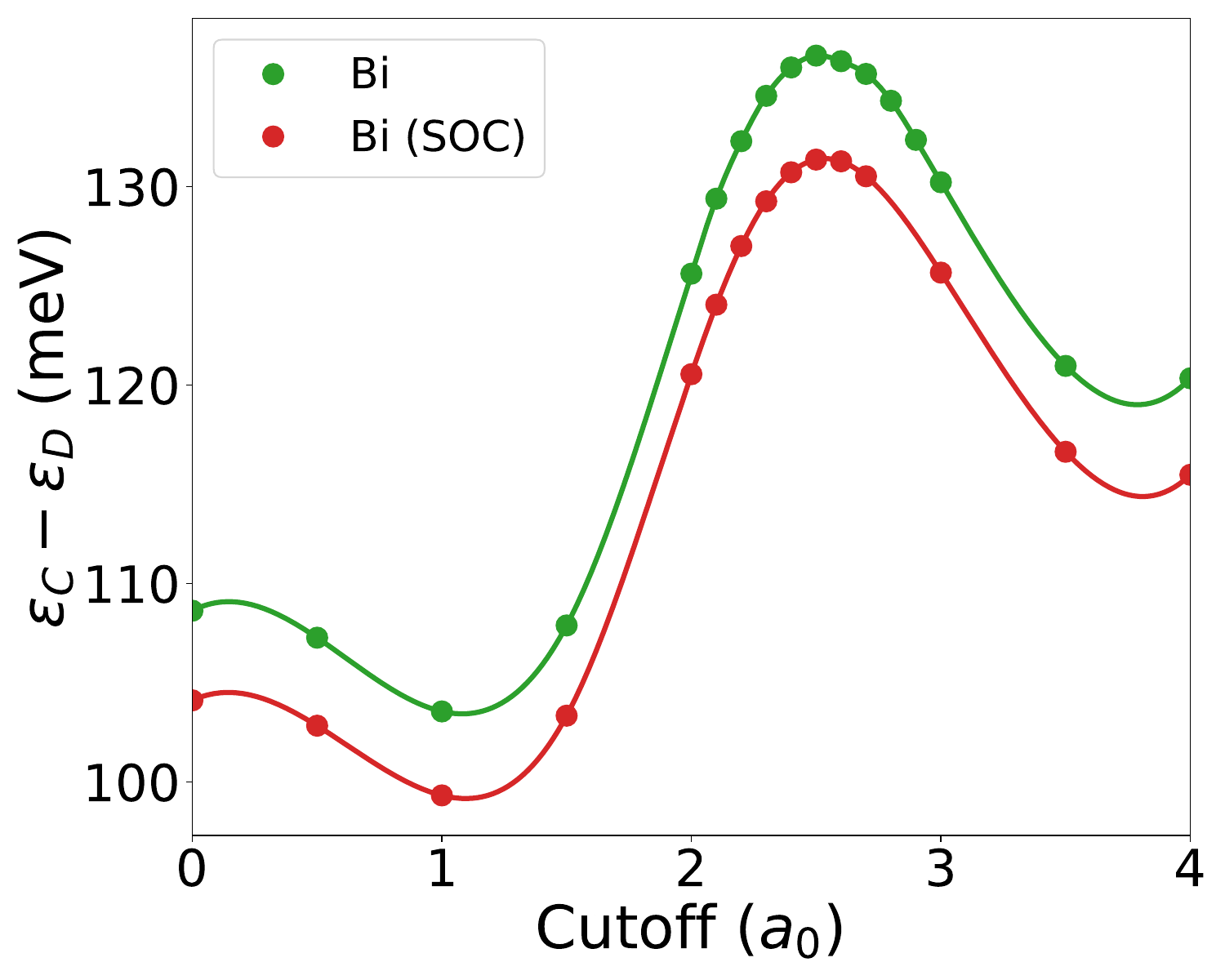}
        \caption{}
        \label{fig:Bi_soc_sweep}
    \end{subfigure}
    \hfill
    \begin{subfigure}{0.49\textwidth}
        \centering
        \includegraphics[width=\textwidth]{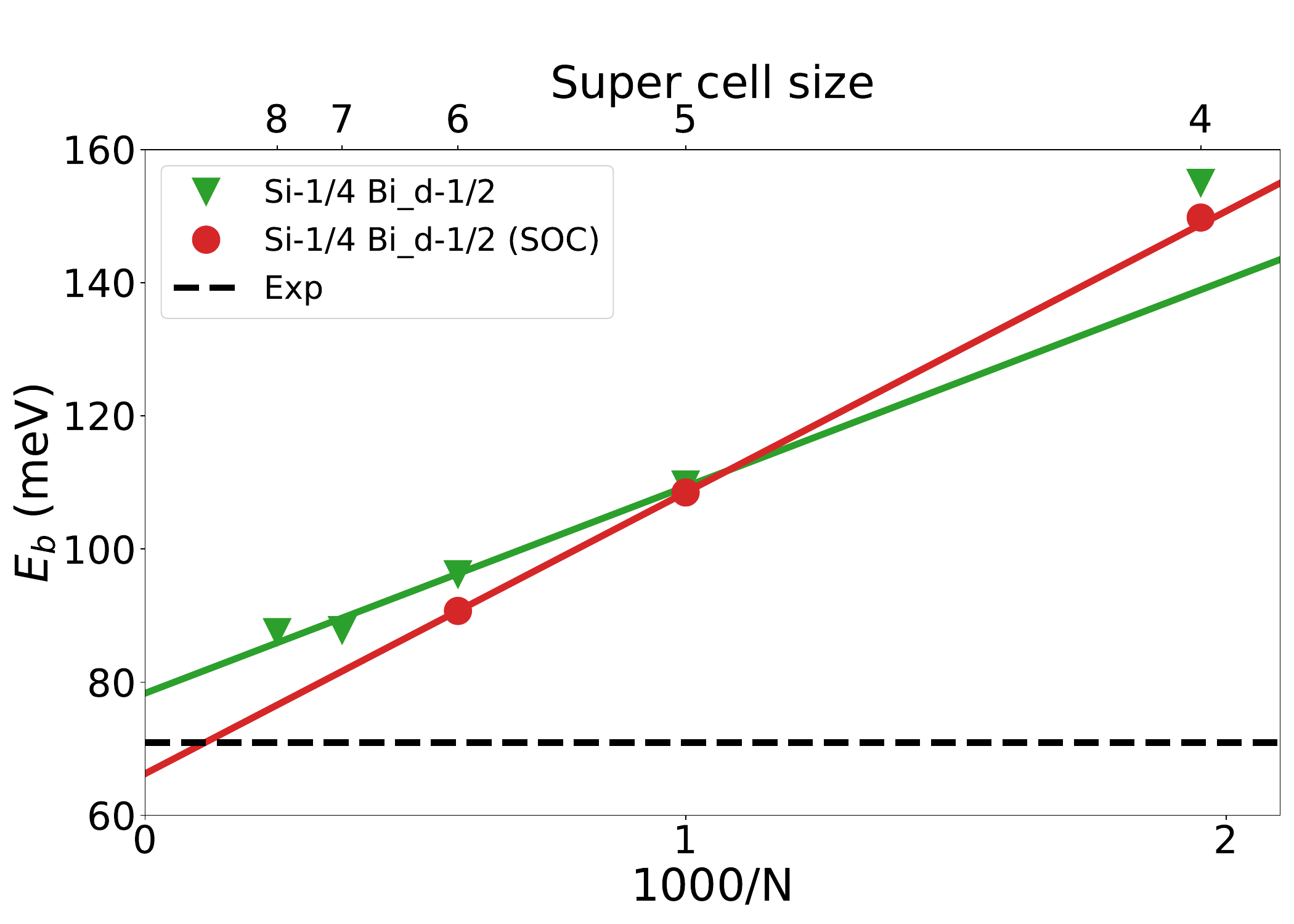}
        \caption{}
        \label{fig:Bi_soc_fit}
    \end{subfigure}
    \caption{(a) The cutoff optimization for bismuth with the SOC enabled. For contrast the cutoff optimization for Bi without SOC is also included. (b) An estimate for the binding energy using the \supercell{5} and \supercell{6} with an without SOC.}
    \label{fig:Bi_soc}
\end{figure}

As we are interested in phenomena at very fine energy scales and are dealing with heavy atoms, we decided to investigate the inclusion of SOC effects. This analysis reveals fundamental physical insights that may have been overlooked in previous computational studies. For Bi, in the scale that we are analyzing of tens of meV, a dramatic reduction is observed in the binding energy upon including SOC (78.52 to 66.25 meV), which brings the computed $E_b$ closer to experiment, as shown in Fig. \ref{fig:Bi_soc_fit}. 
Critically, the HSE calculations by Swift et al. did not include SOC, representing a significant limitation of their approach. Our results demonstrate that SOC reduces binding energies for Bi, suggesting that HSE results would become less accurate if SOC were properly included. 

This analysis highlights a crucial advantage of our \dfthalf approach, the ability to systematically include all relevant physical effects (band gap correction, delocalization correction, relativistic effects) within a unified, computationally tractable framework. While other methods face prohibitive computational costs when including SOC for large supercells, our method enables routine inclusion of these effects. Thus, the inclusion of SOC for bismuth is crucial to obtain more accurate binding energy predictions and simplifies the computational methodology by ensuring adequate occupation of the shallow donor level.

The stepwise correction described above provides  insight into the fundamental limitations of conventional DFT in describing shallow donor states and demonstrates how the DFT-1/2 method systematically addresses these deficiencies. Standard DFT suffers from both band gap underestimation and spurious charge delocalization due to self-interaction errors, causing the shallow donor wavefunction to spread excessively rather than maintaining its proper hydrogenic character. When the bulk DFT-1/2 correction is applied to silicon, the binding energy improves significantly because this correction provides a more accurate band gap and electronic structure of the host material, establishing the proper energy reference for the conduction band minimum. This bulk correction creates a better electronic environment for the shallow state by correcting the host material's electronic structure, though the binding energies still require additional localization corrections to achieve full accuracy. Applying the DFT-1/2 correction directly to the donor atom addresses the remaining self-interaction error due to the spurious charge delocalization in the immediate vicinity of the impurity. This dual-correction approach, bulk band gap correction combined with local impurity self-interaction correction, provides a comprehensive solution to the electronic structure errors inherent in standard DFT. The bulk correction establishes the proper energetic framework through accurate band structure, while the local correction ensures proper charge localization around the donor site. This also explains why the cutoff parameter of the donor atoms is an order of magnitude smaller than the Bohr radius in silicon as it doesn't need to improve the localization of the entire shallow state.
For heavier donor atoms, additional physical effects become important, the inclusion of d-orbitals in the valence description and spin-orbit coupling effects must be considered to achieve quantitative accuracy, providing a complete quantum-mechanical description of the donor atom.



\begin{figure}[h!]
    \centering
    \includegraphics[width=0.8\linewidth]{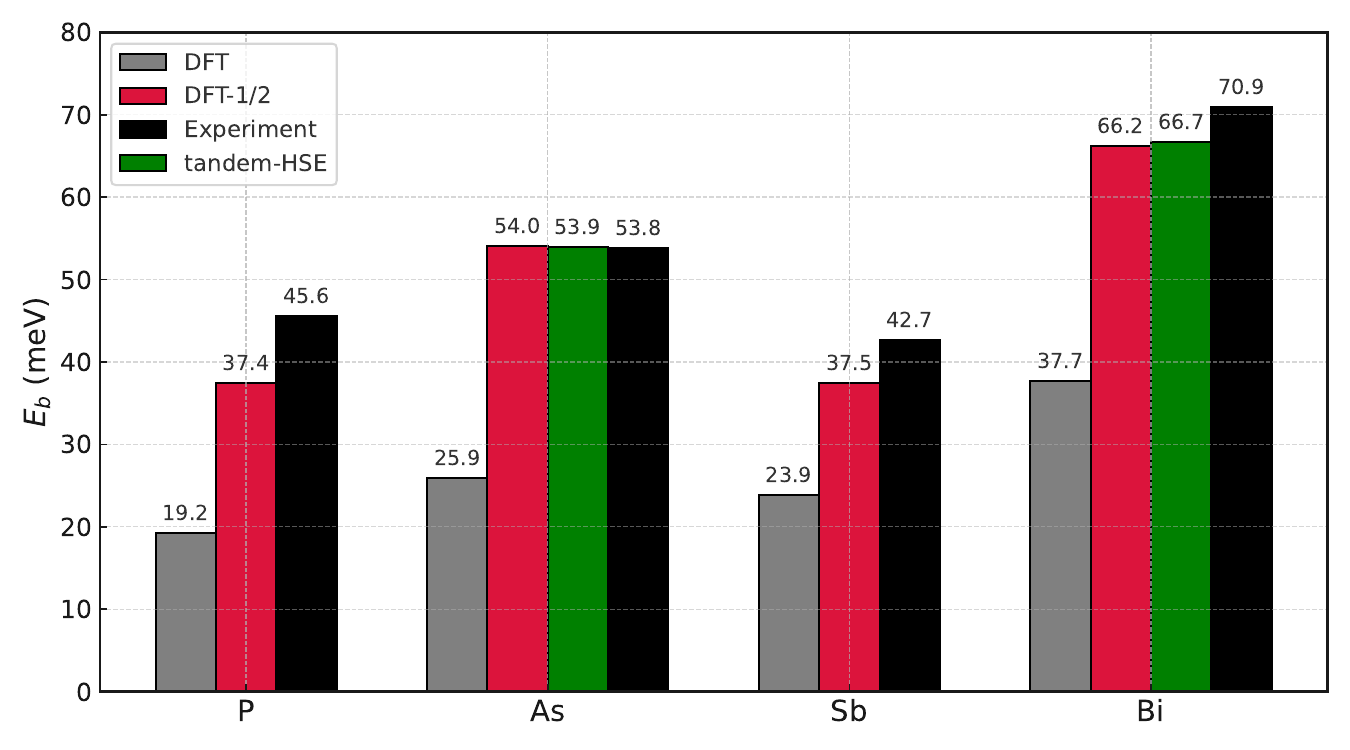}
    \caption{Binding energies $E_b$ (in meV) for group V donors (P, As, Sb, Bi) in silicon, computed using different approaches: standard DFT (gray), DFT-1/2 - best approach (red), and tandem-HSE (green, when available), along with experimental values (black).}
    \label{fig:barchart}
\end{figure}

Figure \ref{fig:barchart} presents a direct comparison of the donor binding energies computed using standard DFT, the DFT-1/2 method (best approach), and the tandem-HSE  (when available), alongside experimental reference values, for all group V impurities (P, As, Sb, Bi) in silicon. The figure illustrates how the DFT-1/2 methodology, while maintaining low computational cost, delivers predictive accuracy on par with that of more demanding approaches.

\begin{table}[H]
\centering
\caption{The binding energy in meV calculated by extrapolating the binding energies of the \supercell{5} up to the \supercell{8} supercell for different levels of correction. The type of correction is indicated in second column called method. The fraction behind each element indicates that a \dfthalf correction is applied. An element with no fraction indicates that an unmodified pseudopotential in the determination of binding energies. Some elements have \_d behind them, this indicates that a pseudopotential with $d$-valence orbitals was used. HSE indicates that the tandem-HSE approach from \cite{Swift2020} was used. The column $a$ contains the slope of the function $E_b(1000/N)=a (1000/N) + E_b$ in meV. The experimental values for the binding energy are obtained from \cite{Saraiva_2015,grimmeiss1982multivalley,pajot2010optical}. The column SC(max) indicates the largest supercell size used in the fit.}
\label{tab:binding_energies}
\begin{tabular}{llrrrr}
\toprule
Donor &            Method &  $E_b$ &  $E_b$(Exp) &   $a$ &  SC(max) \\
\midrule
    \multirow{2}{*}{P}  &                  DFT    &  19.20 & \multirow{2}{*}{45.58}   & 22.07 &        \supercell{7} \\
                        &          \dfthalf Si-1/4 P-1/2     &  37.44 &                          & 24.07 &        \supercell{8} \\ 
                        \midrule
   \multirow{4}{*}{As}  &                 DFT     &  25.93 & \multirow{4}{*}{53.77}   & 27.77 &        \supercell{6} \\
                        &           \dfthalf  Si-1/4 As     &  43.24 &                          & 28.96 &        \supercell{6} \\
                        &        \dfthalf Si-1/4 As-1/2     &  54.04 &                          & 33.34 &        \supercell{8} \\
                       &       tandem-HSE \cite{Swift2020}&  53.9  &                          & 42.1 &        \supercell{6} \\
                        \midrule
   \multirow{3}{*}{Sb}  &                 DFT (Si Sb\_d)  &  23.90 & \multirow{3}{*}{42.71}   & 17.80 &        \supercell{6} \\
                        &       \dfthalf  Si-1/2 Sb-1/2     &  35.73 &                          & 25.12 &       \supercell{7} \\
                        &        \dfthalf Si-1/4 Sb\_d-1/2  &  37.45 &                          & 22.45 &        \supercell{8} \\
                        \midrule
  \multirow{6}{*}{Bi}   &                 DFT &  37.73    & \multirow{6}{*}{70.88}    & 29.67 &        \supercell{6} \\
                        &       \dfthalf Si-1/4 Bi &  58.76    &                           & 30.79 &        \supercell{7} \\
                        &        \dfthalf Si-1/4 Bi-1/2 &  81.46    &                           & 35.98 &          \supercell{8} \\
                        &     \dfthalf  Si-1/4 Bi\_d-1/2 &  78.52   &                           & 31.73 &   \supercell{8} \\
                        & \dfthalf Si-1/4 Bi\_d-1/2 (SOC) &  66.25   &                           & 42.25 &   \supercell{6}\\
                        &    tandem-HSE \cite{Swift2020}&  66.7    &                             & 49.7  &        \supercell{6} \\
\bottomrule
\end{tabular}
\end{table}

\begin{figure}[H]
\centering

\tikzset{
  >=Latex,
  every node/.style = {font=\small},
  box/.style       = {rectangle, rounded corners, draw, thick, align=center, fill=gray!4,
                      inner sep=5pt, minimum width=80mm, text width=80mm},
  titlebox/.style  = {box, fill=gray!12, font=\bfseries, minimum width=110mm, text width=110mm},
  process/.style   = {box},
  decision/.style  = {diamond, draw, thick, aspect=2, align=center, inner sep=1pt, fill=gray!4,
                      text width=38mm},
  line/.style      = {->, thick},
  aux/.style       = {->, thick, dashed} 
}

\begin{tikzpicture}[node distance = 6mm and 18mm]

\node (title) [titlebox] {DFT-1/2 Methodology for Shallow Donor Binding Energies};

\node (s0) [process, below=7mm of title]
  {\textbf{Step 1:Standard DFT calculation and PDOS of pristine bulk}\\Identify VBM orbital character};

\node (s1) [process, below=6mm of s0]
  {\textbf{Step 2:  Standard DFT-1/2 Bulk Correction}\\Correct band-gap underestimation \\
  (consider the corrected potential for the next steps)};

\node (dec) [decision, below=6mm of s1]
  {Impurity \\is a heavy atom?};

\draw[line] (dec.west) -- ++(-22mm,0) node[above, align=center, xshift=-1mm]
  {\footnotesize No:\\\footnotesize standard PP} coordinate (L);
\draw[line] (dec.east) -- ++( 22mm,0) node[above, align=center, xshift= 1mm]
  {\footnotesize Yes:\\\footnotesize include $d$-valence/SOC} coordinate (R);

\node (s2) [process, below=6mm of dec]
  {\textbf{Step 3: PDOS of doped material}\\Identify donor orbital character};

\node (s3) [process, below=6mm of s2]
  {\textbf{Step 4: Donor Atom Correction}\\
   Apply \dfthalf correction to the donor \emph{contributing} orbital\\
   Optimize cutoff $\Rightarrow$ maximize $\varepsilon_{C}-\varepsilon_{D}$\\
   Correct donor localization};

\node (s4) [process, below=6mm of s3]
  {\textbf{Step 5: Supercell Calculations}\\
   Multiple sizes ($5\!\times\!5\!\times\!5$ to $8\!\times\!8\!\times\!8$)};

\node (s5) [process, below=6mm of s4]
  {\textbf{Step 6: Obtaining Impurity Binding Energy}\\
   $E_b \!=\! \varepsilon_{CBM}^{\Gamma} - \varepsilon_{D}^{\Gamma} + e\,\Delta V$\\
   Linear fit vs.\ $1/N \Rightarrow$ intercept at $1/N \to 0$};

\draw[line] (title) -- (s0);
\draw[line] (s0) -- (s1);
\draw[line] (s1) -- (dec);

\draw[line] (dec.south) -- (s2.north);
\draw[line] (L) |- (s2.west);
\draw[line] (R) |- (s2.east);

\draw[line] (s2) -- (s3);
\draw[line] (s3) -- (s4);
\draw[line] (s4) -- (s5);

\end{tikzpicture}

\caption{Workflow for shallow donor binding energies using DFT-1/2.}
\label{fig:dft12_workflow_predec}
\end{figure}
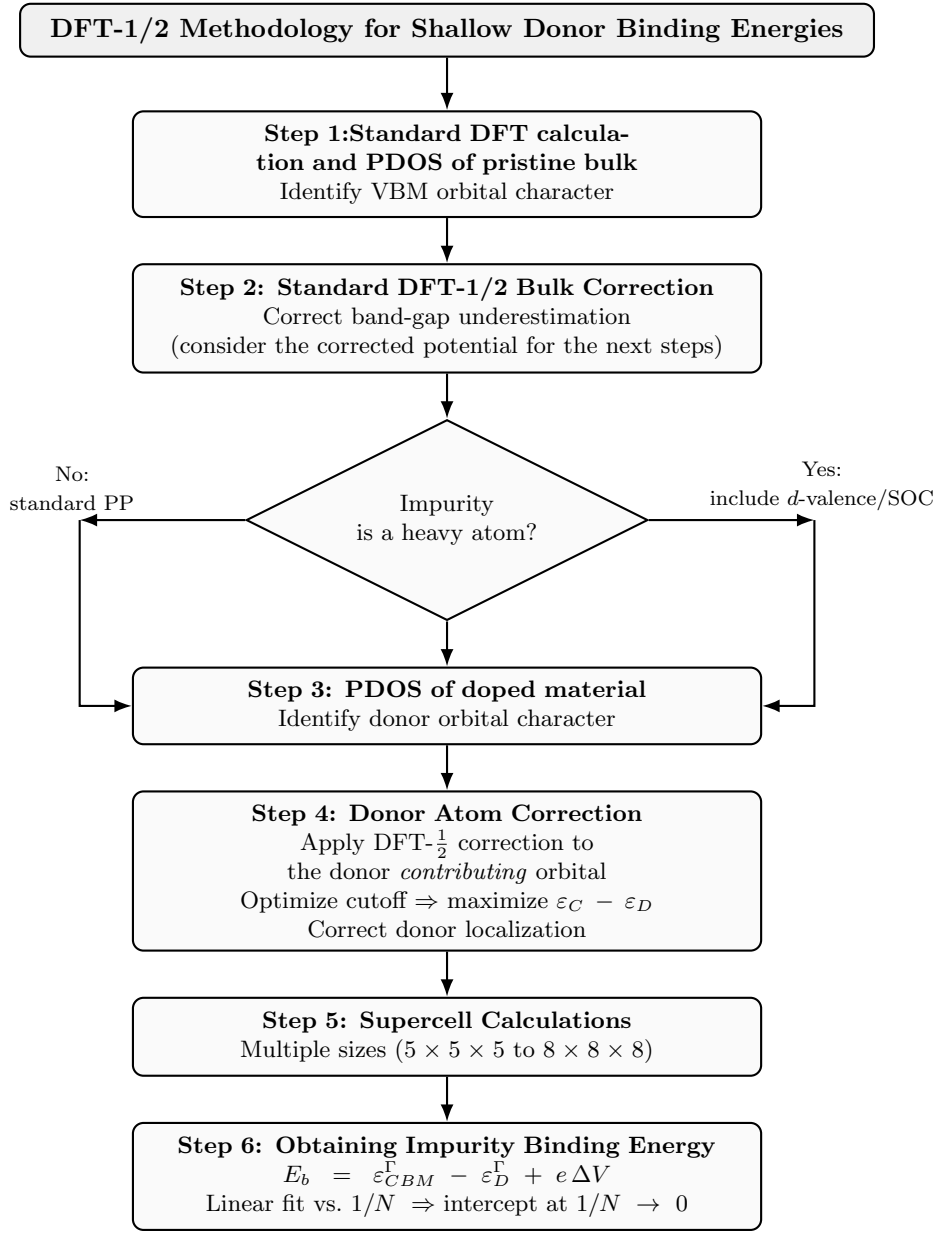

Figure \ref{fig:dft12_workflow_predec} outlines the workflow adopted in this study for calculating shallow donor binding energies using the DFT-1/2 approach. The procedure begins with standard DFT calculations on pristine material to characterize the host material, in the sequence the DFT-1/2 method is applied to correct its band gap. If the impurity is a heavy atom (e.g., Sb, Bi), $d$-valence orbitals and SOC are included. Subsequently, PDOS analysis is performed to identify the donor orbital character. The DFT-1/2 donor correction is then applied to the relevant donor orbital, with the cutoff parameter optimized to maximize the separation between the shallow level and the nearest conduction state. Finally, supercells of increasing size (up to 8×8×8, some with embedded method for relaxation) are used to extrapolate the binding energy to the dilute limit. This structured approach enables accurate and transferable predictions while remaining computationally efficient.

\section{Conclusion}

In this study, a comprehensive investigation of the effectiveness of the \dfthalf correction for accurately determining the binding energy for group V shallow donors in silicon is presented. Our approach systematically addresses the inherent limitations of standard DFT, such as band gap underestimation and electron delocalization errors, while crucially maintaining a high level of computational efficiency. We showed that it is essential to correct both the bulk silicon atoms and the donor atom. By doing so we found remarkable improvements upon the DFT results, which are known to severely underestimate binding energies. For arsenic, the calculated binding energy of 54.04 meV is in excellent agreement with the experimental value of 53.77 meV, differing by only 0.27 meV. This result is highly competitive with the tandem-HSE value of 53.9 meV reported in the literature. For the bismuth donor, initial \dfthalf calculations without SOC overestimated the experimental value (78.52 meV versus 70.88 meV). However, by including SOC in our calculations, the binding energy significantly decreased to 66.25 meV. This SOC-corrected value is closer to the experimental value of 70.88 meV, differing by approximately 4.63 meV. For the other Group V donors, antimony and phosphorus, the \dfthalf method also provided substantial improvements. For antimony, the binding energy prediction of 37.45 meV is within approximately 5 meV of the experimental value of 42.71 meV. For phosphorus, while presenting a great improvement with respect to the DFT, the result of 37.44 meV deviates by approximately 8 meV from the experimental value of 45.58 meV.

Beyond its enhanced accuracy, the \dfthalf method offers a significant practical advantage, as it is less complicated and less computationally demanding than other methods. While HSE calculations face limitations in supercell size due to large computational costs, and needs a tandem approach, i.e., requires combining results from different functionals and pseudopotential treatments, our methodology employs a single, unified functional consistently across all supercell sizes, eliminating the need for complex extrapolation procedures between different levels of theory. Therefore, the \dfthalf allowed us to perform calculations on supercells of up to 4096 atoms, crucial for capturing the extended nature of shallow defect wavefunctions. Moreover, this developed \dfthalf methodology can be generalized effectively to other shallow impurities in semiconductors. Furthermore, spin-orbit coupling can be seamlessly incorporated within the DFT-$\frac{1}{2}$ framework with the same computational overhead as in standard DFT calculations, enabling the treatment of relativistic effects that are often neglected in more demanding methods due to prohibitive costs. 

In summary, the \dfthalf method represents a significant step forward in the ab initio simulation of  shallow impurities in semiconductors, providing highly accurate binding energies with a favorable computational cost, and offering a robust platform for further investigations into the electronic structure of these critical quantum systems.

\section*{Data availability}
The data supporting this study are available in the NOMAD repository at \url{https://doi.org/710.17172/NOMAD/2025.09.23-1}.

\section*{Code availability}
The code used to determine the \dfthalf cutoffs is available at \url{https://github.com/JoshuaClaes/defectonehalf}.

\printbibliography

\section*{Acknowledgements} 
This work was performed in part using HPC resources from the VSC (Flemish Supercomputer Center) and the HPC infrastructure of the University of Antwerp (CalcUA), both funded by the FWO-Vlaanderen and the Flemish Government department EWI (Economie, Wetenschap \& Innovatie).

\section*{Funding statement}
This work was supported by the FWO (Research Foundation-Flanders), project G0D1721N.  L.K.T thanks the  National Council for Scientific and Technological Development (CNPq), under grants 316081/2023-0 and 444563/2024-5, and the Sao Paulo Research Foundation 
(FAPESP) under grant 2006/05858-0.\\
The funders played no role in study design, data collection, analysis and interpretation of data, or the writing of this manuscript.

\section*{Author contributions}
J.C. performed all simulations, developed the workflow, and made substantial contributions to writing and formatting the manuscript. J.C., B.P., and D.L. conceptualized the study. B.P. and D.L. also provided guidance throughout the project and contributed to writing. M.M. offered advisory input, assisted in writing, and provided additional insight and expertise on the DFT-1/2 method. L.T. contributed substantially to writing and formatting, provided advisory support, and contributed additional insight and expertise on the DFT-1/2 method. All authors discussed the results and reviewed the manuscript.

\section*{Competing interests}
The authors declare no competing interests.

\pagebreak

\begin{appendices}

\section{The energy cutoff}
\begin{figure}[H]
    \centering
    \begin{subfigure}{0.49\textwidth}
        \includegraphics[width=\linewidth]{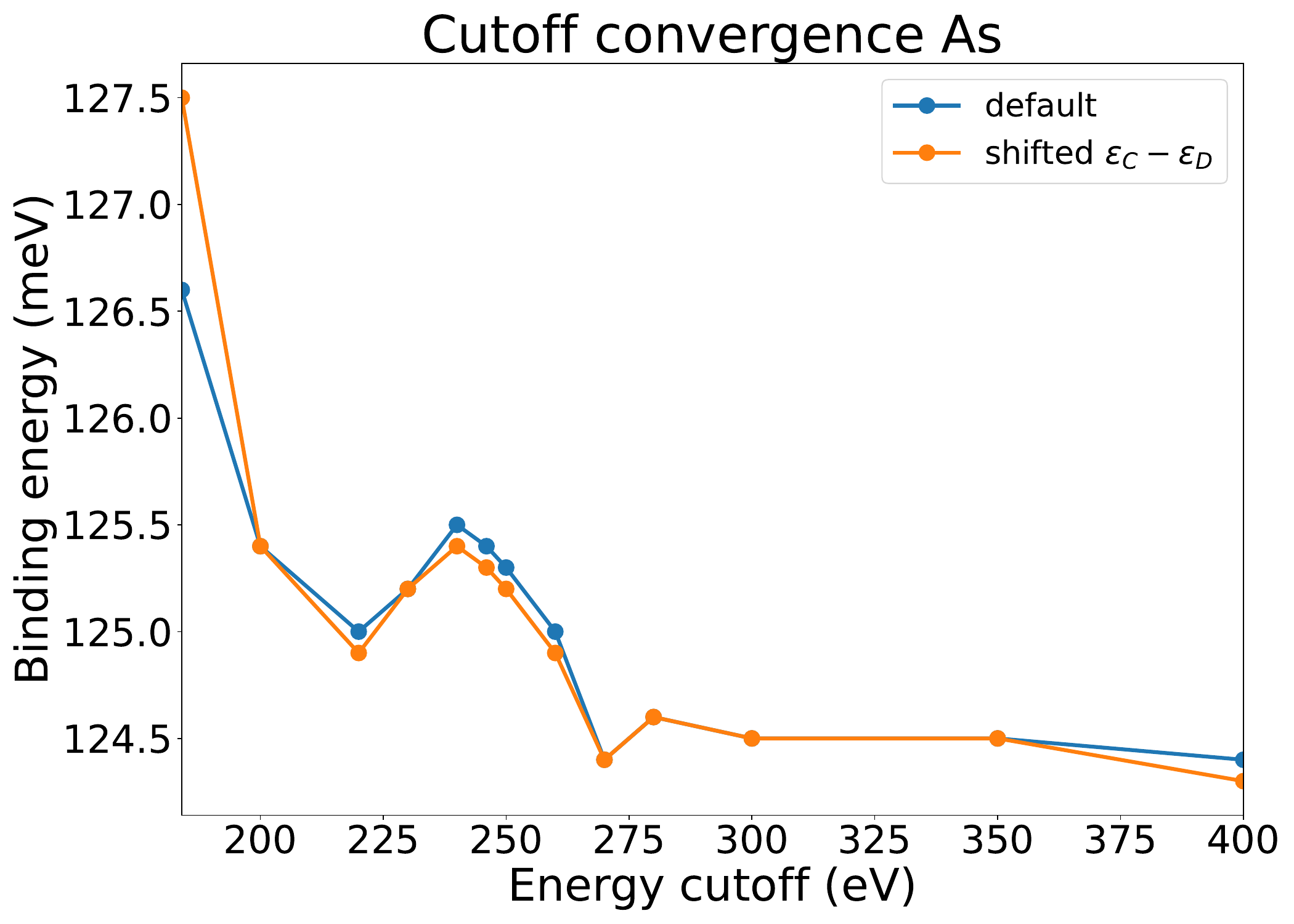}
        \caption{}
        \label{fig:encut_conv_As}
    \end{subfigure}
        \begin{subfigure}{0.49\textwidth}
        \includegraphics[width=\linewidth]{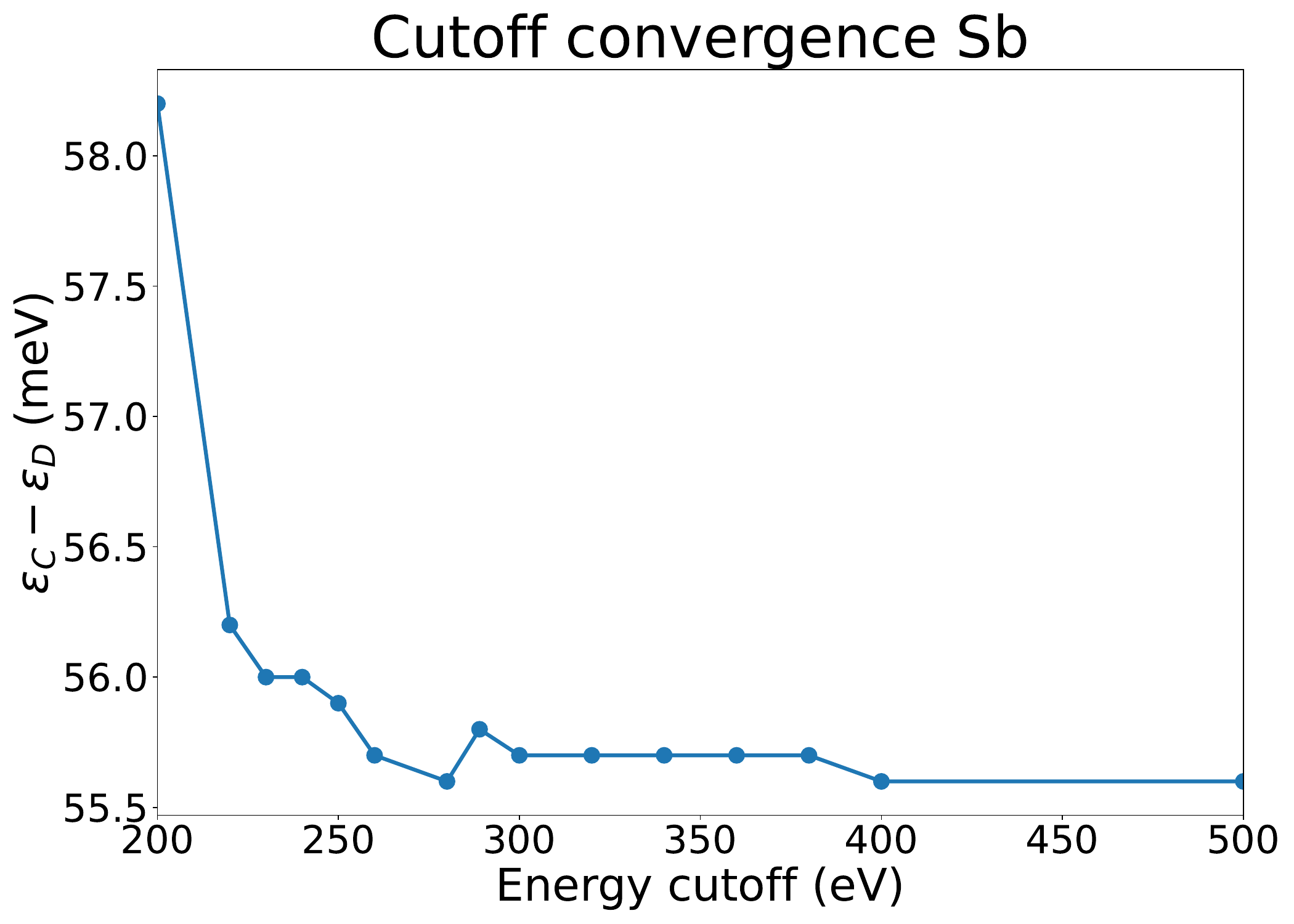}
        \caption{}
        \label{fig:encut_conv_Sb}
    \end{subfigure}
    \caption{The figure shows the convergence of the separation $\varepsilon_C - \varepsilon_D$ for arsenic (a) and antimony (b) as a function of the cutoff parameter. For arsenic the binding energy is also shown and the separation was shift such that the binding energy and the separation have the same value at a cutoff of 246 eV. This is done to show that the separation is a good cutoff criteria.}
    \label{fig:shallow_encut_conv}
\end{figure}

In the main text, we use the energy cutoff values shown in Tab. \ref{tab:shallow_encuts}. In most cases, the energy cutoff was set to the highest ENMAX value from the POTCAR file for the system. Since the binding energy is a very small value, a low energy cutoff can significantly affect the reliability of the calculation. Figure \ref{fig:encut_conv_As} examines the convergence of the binding energy as a function of the energy cutoff. The chosen energy cutoff of 246 eV differs by only approximately $1$ meV from the largest cutoff value of $400$ eV.

The same figure also compares the convergence of the binding energy with that of the separation $\varepsilon_C - \varepsilon_D$, which is the gap between the shallow level and the conduction band of the same spin right above. The separation $\varepsilon_C - \varepsilon_D$ in the doped system appears to serve as a convergence criterion. This is convenient because it means we don’t need to calculate the pristine system at each tested energy cutoff.

Fig. \ref{fig:encut_conv_Sb} checks the energy cutoff convergence for antimony. For this system, $\varepsilon_C - \varepsilon_D$ seems to converge well before the ENMAX value of 289 eV.

\section{\dfthalf cutoff parameter convergence} \label{Apx:dfthalf_cutoff_conv}
\begin{figure}[H]
    \centering
    \begin{subfigure}{0.48\textwidth}
        \includegraphics[width=\linewidth]{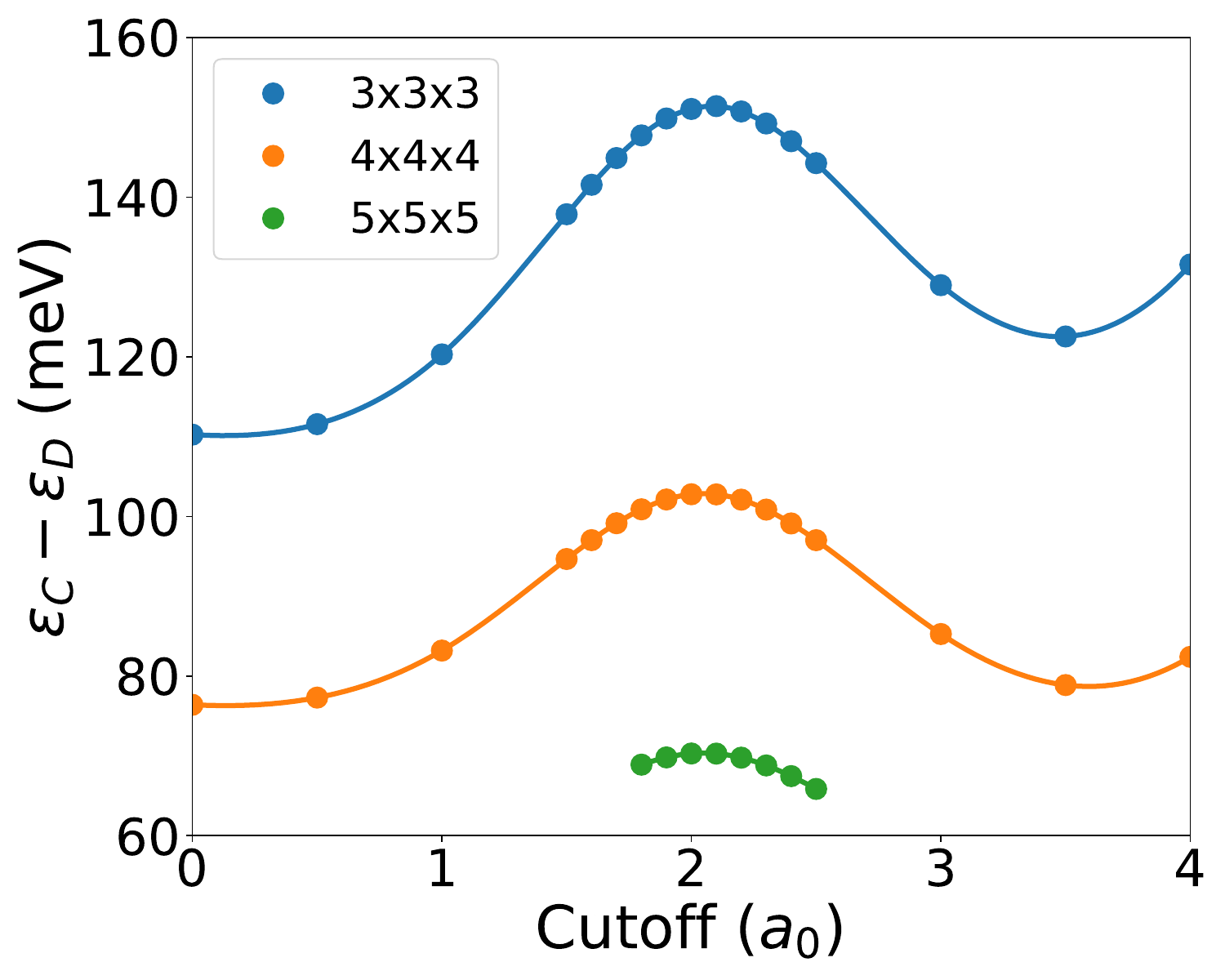}
        \caption{As}
        \label{fig:rc_conv_As}
    \end{subfigure}
    \begin{subfigure}{0.48\textwidth}
        \includegraphics[width=\linewidth]{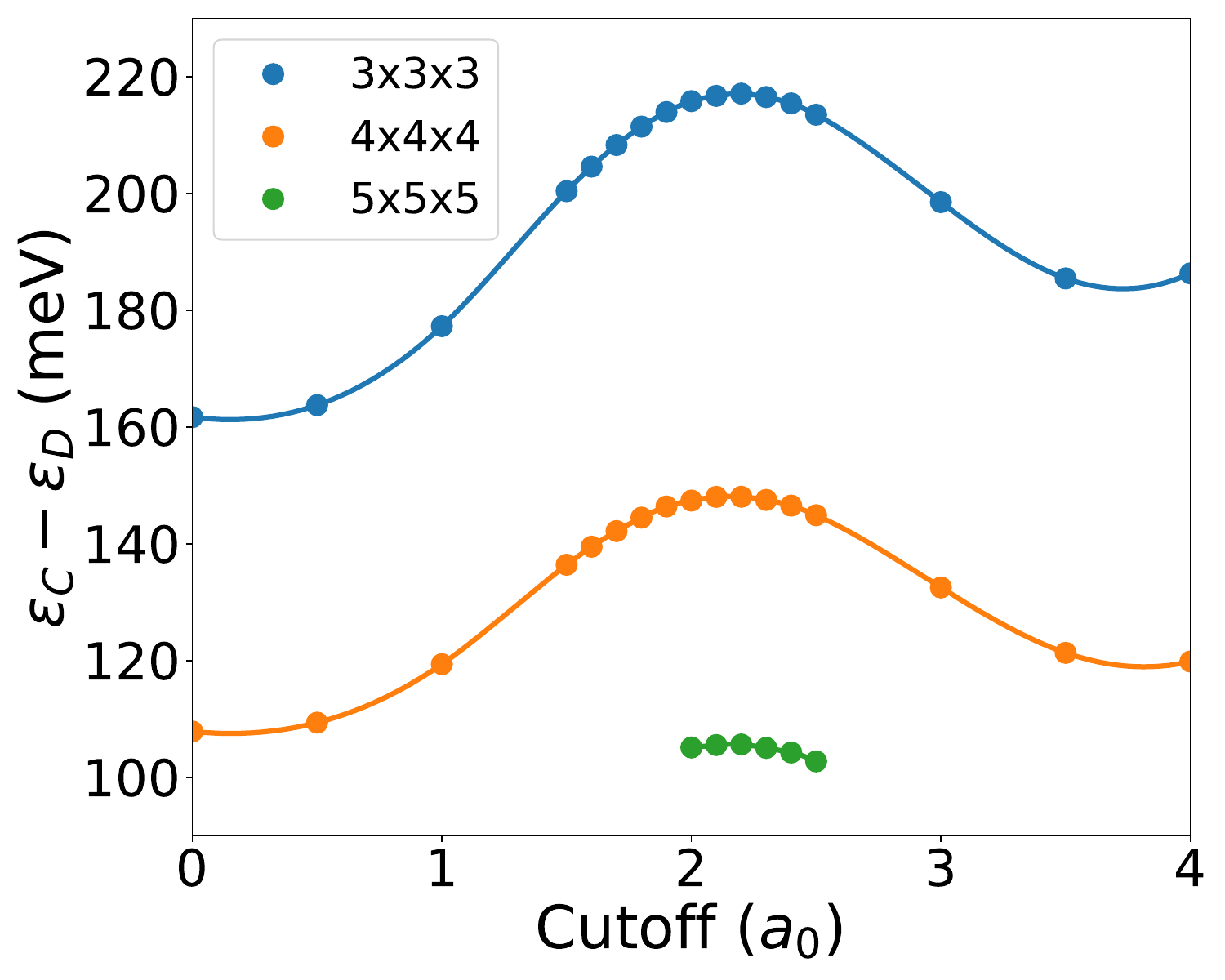}
        \caption{Bi}
        \label{fig:rc_conv_Bi}
    \end{subfigure}
    \caption{The extremization of $\varepsilon_C-\varepsilon_D$  of the same spin for As and Bi for super cell sizes ranging form \supercell{3} to \supercell{5}. These calculations where done without valence $d$-orbitals.}
    \label{fig:shallow_rc_conv}
\end{figure}

To maintain the efficiency of the shallow defect calculation procedure, it would be ideal if the cutoff radius, $r_c$, could be determined on a small supercell. This would minimize the number of calculations required for the large supercell. We therefore checked the convergence of the cutoff parameter for different supercell sizes, as shown in Fig. \ref{fig:shallow_rc_conv} and Tab. \ref{tab:shallow_rc_conv}. The cutoff parameter seems to converge rather quickly, and a supercell size of $4\times4\times4$ is sufficient to correctly determine this parameter.

\begin{table}
    \centering
    \begin{tabular}{ccccc}
    \toprule
    Supercell& \multicolumn{2}{c}{As} & \multicolumn{2}{c}{Bi} \\
        \cmidrule(lr){2-3} \cmidrule(lr){4-5}
         & $r_c$ & $E_b$ & $r_c$  & $E_b$\\ \midrule
        \supercell{3} & 2.1 & 151.4 & 2.2 & 217.0\\
        \supercell{4} & 2.0 & 102.8 & 2.2 & 148.1\\
        \supercell{5} & 2.0 &  70.3 & 2.2 & 105.6\\
        \bottomrule
    \end{tabular}
    \caption{The optimal cutoff parameter in $a_0$ and binding energy in meV for different supercell sizes determined from the cutoff sweep of Fig. \ref{fig:shallow_rc_conv}. }
    \label{tab:shallow_rc_conv}
\end{table}

\begin{figure}[H]
    \centering
    \begin{subfigure}{0.48\textwidth}
        \includegraphics[width=\linewidth]{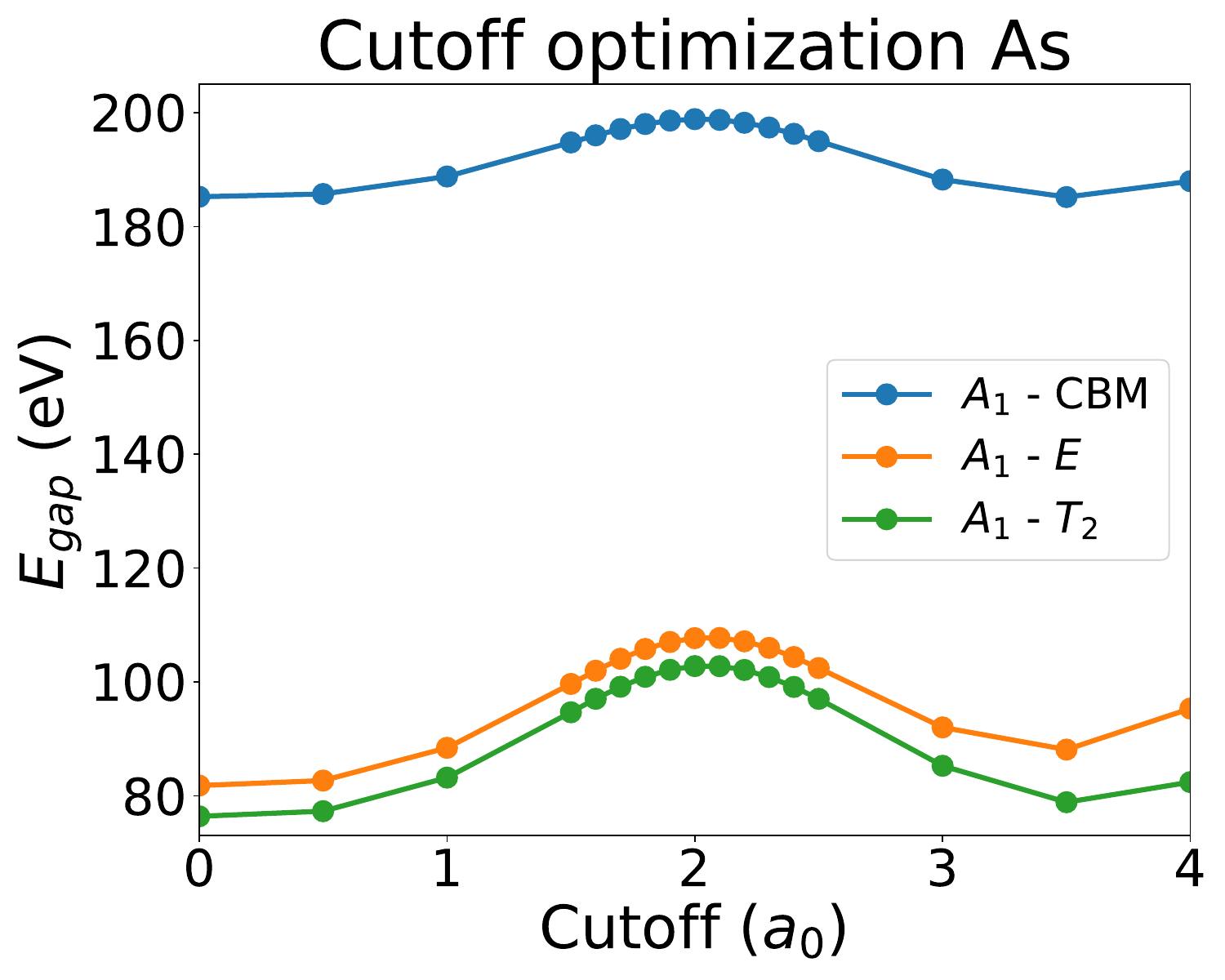}
        \caption{As}
        \label{}
    \end{subfigure}
    \begin{subfigure}{0.48\textwidth}
        \includegraphics[width=\linewidth]{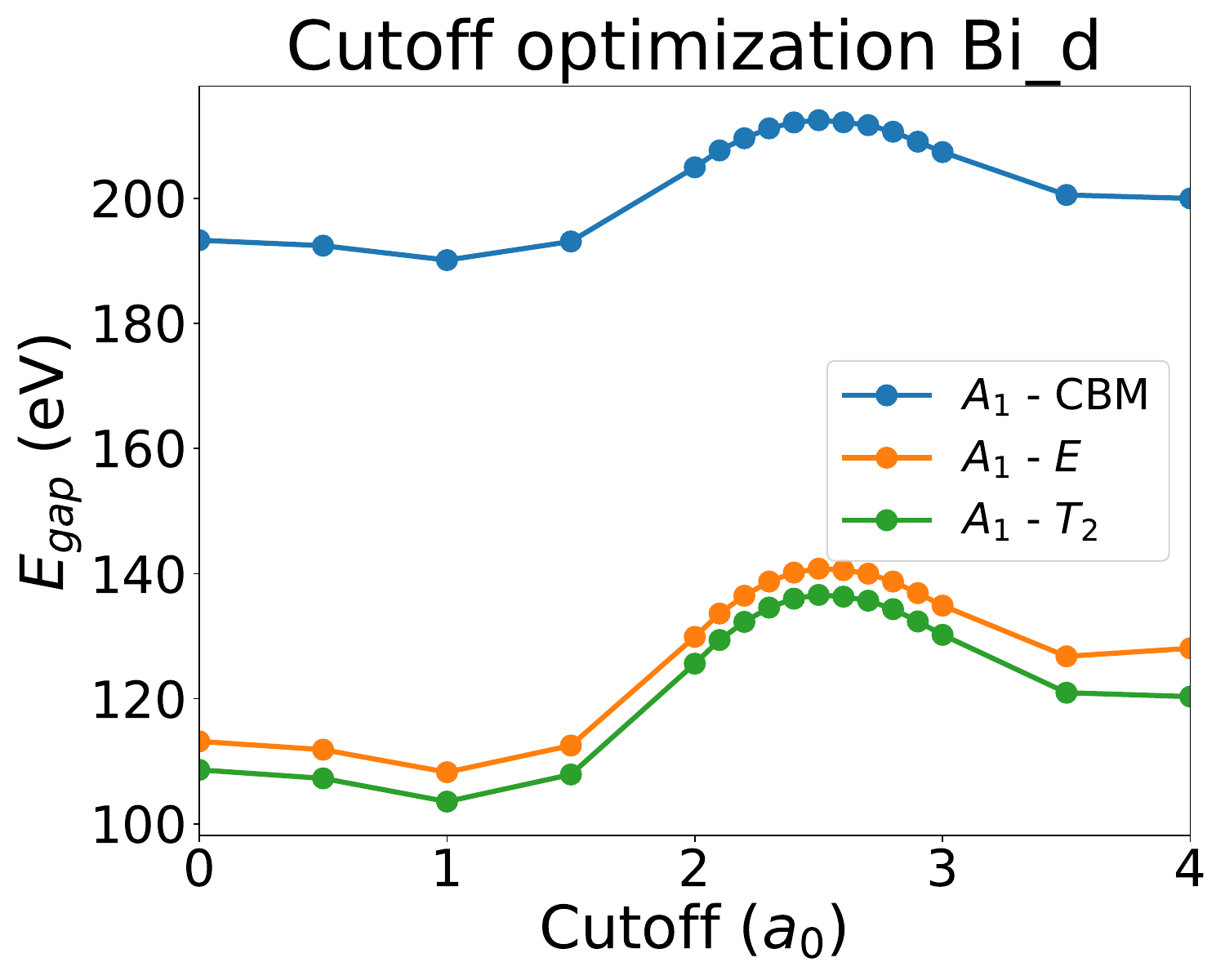}
        \caption{Bi\_d}
        \label{}
    \end{subfigure}
    \caption{The cutoff optimization for As and Bi\_d for the gap between $A_1$ and the other shallow levels along with the gap between  $A_1$ and the CBM}
    \label{fig:shallow_different_gaps}
\end{figure}

A challenge when dealing with shallow defects lies in selecting the appropriate gap to optimize. Typically, bulk calculations focus on maximizing the gap between the highest occupied and lowest unoccupied levels, the band gap. However, in our calculations, directly maximizing this would optimize the gap between the occupied $A_1$ shallow level and the unoccupied $T_2$ shallow level. Instead, we are interested in the binding energy, which represents the gap between the $A_1$ level and the Conduction Band Minimum (CBM)\footnote{If the shallow level $A_1$ is at index $n$, we need to optimize the gap between $\varepsilon_n$ and $\varepsilon_{n+6}$ to obtain the gap between $A_1$ and the CBM}. Figure \ref{fig:shallow_different_gaps} illustrates the various gaps between the $A_1$ shallow level and other shallow levels, as well as the gap with the CBM, during the optimization process for the cutoff parameter. We observed that optimizing for any of these gaps leads to cutoff parameters within $0.1$ $a_0$ of each other. For the \supercell{4}, an error of $0.1$ $a_0$ in the cutoff parameter results in an error of at most $\sim 0.4$ meV in the seperation $\varepsilon_C - \varepsilon_D$ and this error decreases with supercell size. For example for the \supercell{5} supercell this error is at most $\sim 0.1$ meV. Given the close proximity of all gaps shown in Fig. \ref{fig:shallow_different_gaps}, we selected the $A_1 - T_2$ gap to determine the cutoff parameter, maintaining consistency with the typical bulk method.

\section{Convergence of binding energy calculations} \label{apx:sc_conv_Eb}
\begin{figure}[H]
\begin{subfigure}{0.52\textwidth}
    \centering
    \includegraphics[width=\textwidth]{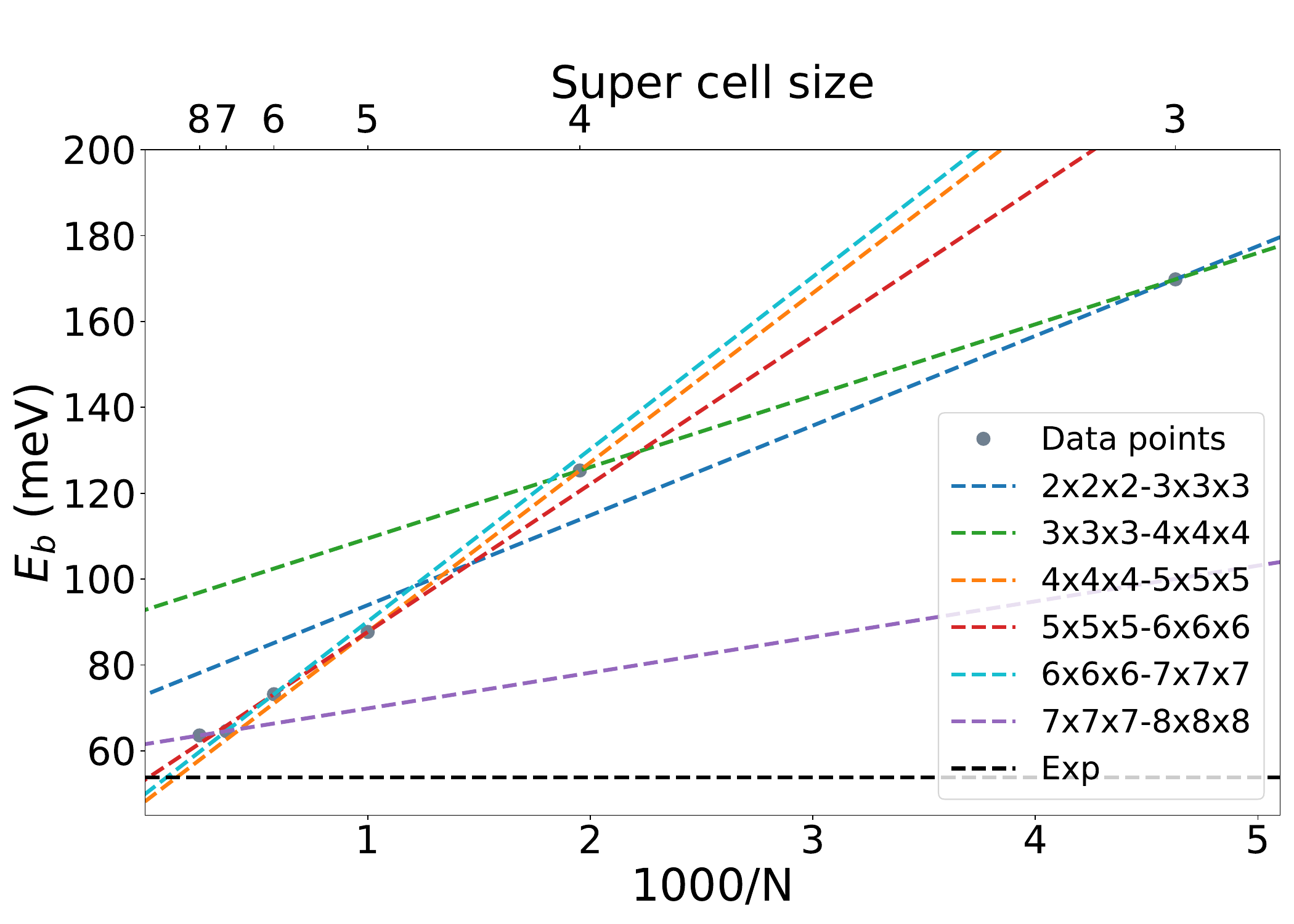}
    \caption{}
    \label{fig:As_Eb_conv_fit_2}
\end{subfigure}
\begin{subfigure}{0.47\textwidth}
    \centering
    \includegraphics[width=\textwidth]{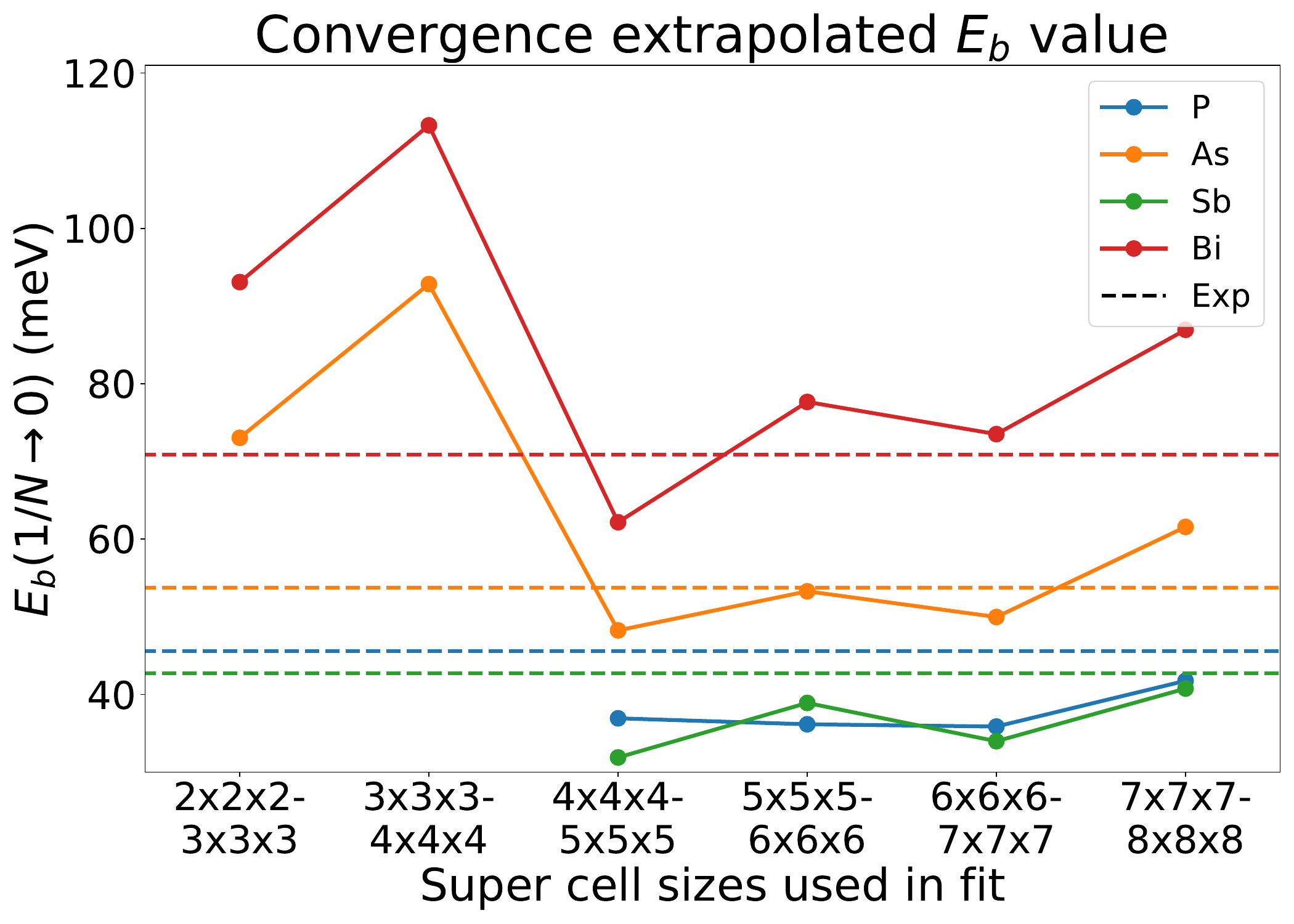}
    \caption{}
    \label{fig:Eb_conv_fit_2}
\end{subfigure}
\begin{subfigure}{0.52\textwidth}
    \centering
    \includegraphics[width=\textwidth]{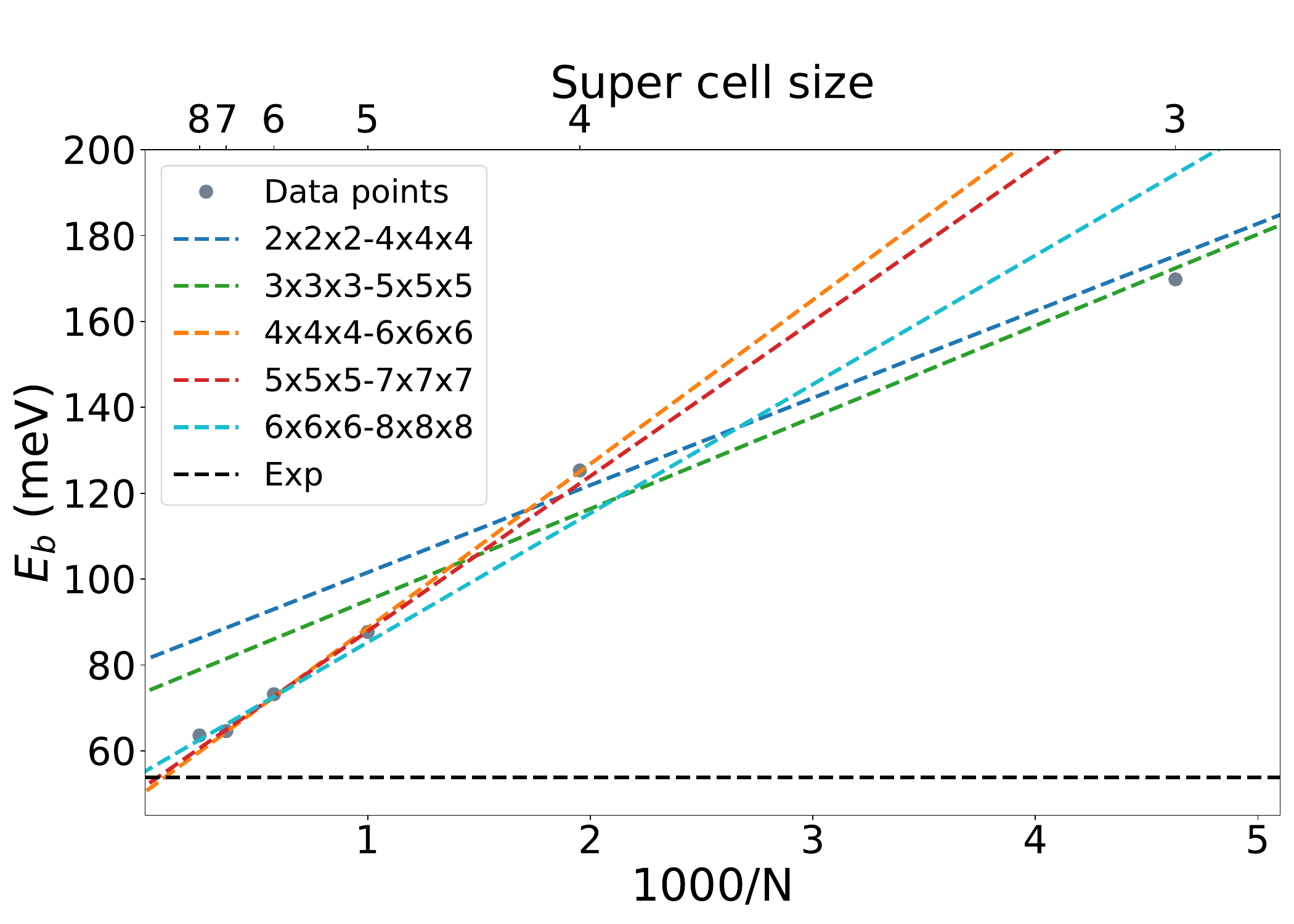}
    \caption{}
    \label{}
\end{subfigure}
\begin{subfigure}{0.47\textwidth}
    \centering
    \includegraphics[width=\textwidth]{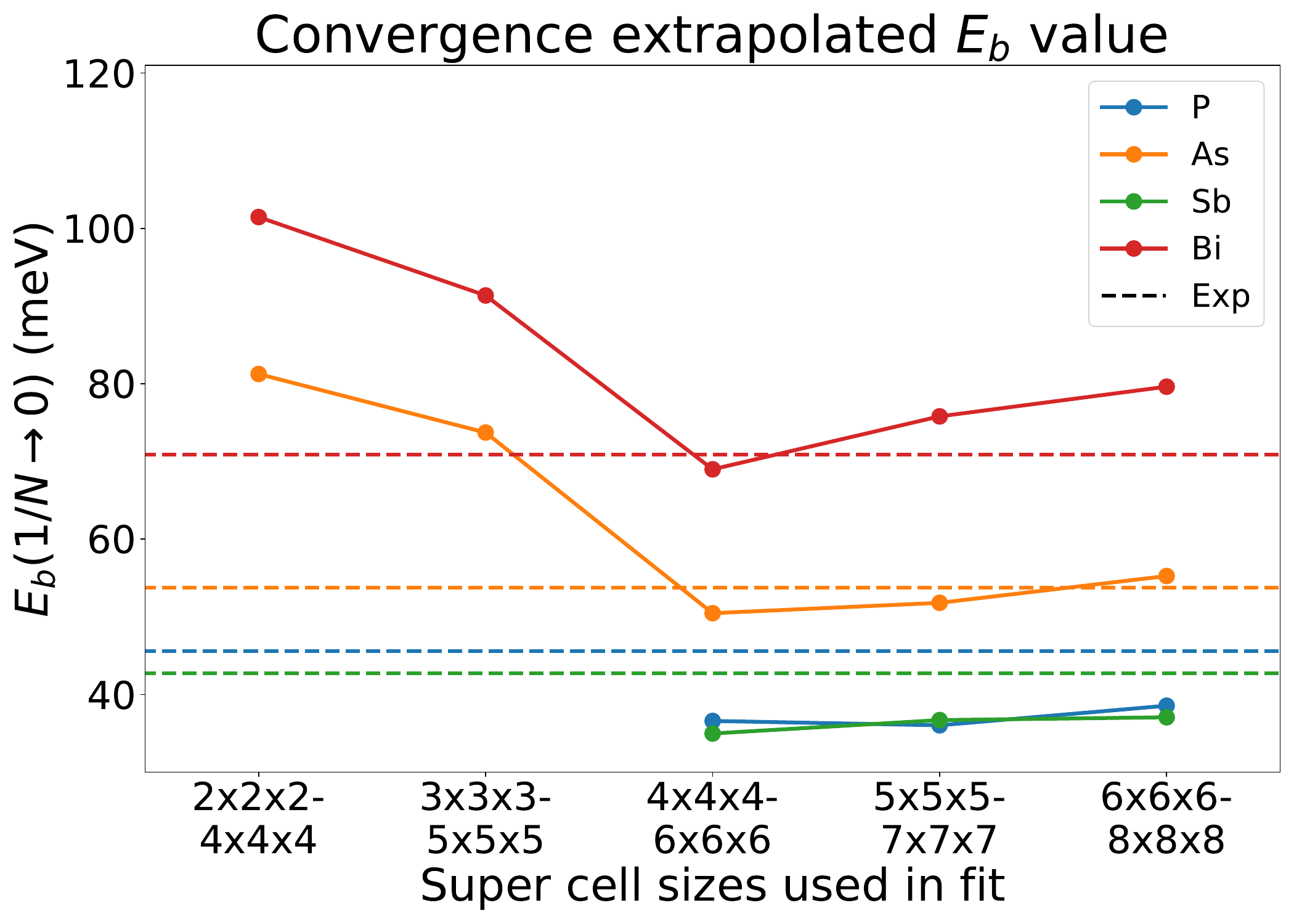}
    \caption{}
    \label{}
\end{subfigure}
\caption{The fit for the binding energy made with sets of 2 (a-b) and 3 (c-d) supercell sizes ranging from \supercell{2} to \supercell{8}. In (a) the fits for As is shown along side the binding energy for each supercell size. In (b) the extrapolated binding energy ($E_b(1/N\rightarrow0)$) is plotted for all fits and all defects. The experimental value, as given in Tab. \ref{tab:binding_energies}, is plotted in the same color.}
\label{fig:Eb_conv_supercell}
\end{figure}

To efficiently apply the \dfthalf method for binding energies, it is important to know which supercell sizes should be included in the fit predict the binding accuracy accurately. In Fig. \ref{fig:Eb_conv_supercell} the fits and predict binding energies are shown for different supercell sizes. In seems like the binding energy converges for fits with supercell sizes of \supercell{4} or higher. However, a closer look at Fig. \ref{fig:shallow_lin_fits} and \ref{fig:Bi_soc_fit} shows that the binding energy of the \supercell{4} still significantly deviates from the actual fits. We therefore chose to ignore this point in all our fits and started fitting with the \supercell{5} supercells and larger. But we do note that the \supercell{4} and \supercell{5} can already yield a reasonable prediction for the final value.

If we focus on the fits of the two largest supercell size i.e. \supercell{7} and \supercell{8} in Fig. \ref{fig:As_Eb_conv_fit_2} and \ref{fig:Eb_conv_fit_2}, we notice that the prediction is off. This is likely due to small numerical error for the predicted binding energy, to which the larger supercells are more sensitive, combined with a small difference in $1000/N$ value resulting incorrect prediction for the final binding energy. Therefore it is important to also include the fits of the smaller supercell sizes of \supercell{5} and \supercell{6}.

\section{Literature HSE result for Phosphorus}\label{apx:HSE_P_literature}
It remains unclear why the \dfthalf results for the P dopant exhibit lower accuracy compared to the other dopants. We note that there is a work in literature by Ma et al. \cite{Ma2022} which determines the binding energy with the tandem HSE approach for P, thereby providing DFT calculations against which we can compare our results. When comparing our DFT binding energy with those of \cite{Ma2022}, we observe that our binding energy of $19.20$ meV is significantly lower than their prediction of $32.5$ meV. 

Had our DFT result been closer to their value, it is likely that our final \dfthalf prediction would better align with experimental data. All binding energies for specific supercell sizes and intermediate results from these calculations can be found in Appendix \ref{apx:intermediate_results}.

However, we suspect that the predicted values by Ma et al. are inaccurate. Given that their input files are publicly available \cite{Ma2022_data}, we attempted to reproduce the binding energy for the \supercell{4} and \supercell{5} supercells. Despite the availability of their input files, we were unable to recreate their reported binding energies of $93.21$ meV and $64.05$ meV, consistently obtaining binding energies of $70.70$ meV and $42.20$ meV for the \supercell{4} and \supercell{5} supercells, respectively. This aligns with our result of $65.7$ meV and $41.2$ meV, but not with the published values. Upon closer inspection, our values for $\varepsilon_C$ and $\varepsilon_D$ (see Tab. \ref{tab:int_res_structure_MA}) are in agreement\footnote{They differ by less than $1$ meV, consistent with the precision of the results presented in \cite{Ma2022}.} with those found in the supplemental material of \cite{Ma2022} (also shown in Tab. \ref{tab:int_res_MA}).

However, the electrostatic energy shift $e\Delta V$ differs significantly between our calculations. We find a value of $17.5$ meV for the \supercell{4}, while they report $45$ meV, which roughly 50\% of their calculated binding energy. We determined $e\Delta V$ using both the histogram method (as described in the main text) and the tetrahedron method \cite{H_CdO_Amini}, both yielding the aforementioned result of $17.5$ meV\footnote{This result is anticipated, as the methods are equivalent}. Additionally, we employed the method of Freysoldt et al. \cite{Freysoldt2009, Freysoldt2011}, the methodology used in \cite{Ma2022}, which resulted in a value of $12.2$ meV.

Furthermore, comparing the values of $e \Delta V$ for arsenic and bismuth with those reported in the article by Swift et al. \cite{Swift2020}, we find that our values differ by at most $4$ meV, attributable to slight variations in the computational setup and the use of the method by Freysoldt for calcualting $e\Delta V$ by Swift et al.

Given these discrepancies, we suspect that Ma et al. may have incorrectly calculated the values for $e \Delta V$, and that their agreement with experiment is coincidental\footnote{We have only examined the VASP inputs and not the CP2k inputs used in this work. However, both methodologies yield comparable values for $e \Delta V$}. Therefore, the predicted value for $E_b$ using the tandem HSE approach is likely overestimated, suggesting that the tandem HSE predictions aligns more closely with our \dfthalf results.

\begin{table}[H]
    \centering
    \begin{tabular}{lccc}
    \toprule
    Supercell & $\varepsilon_\Gamma^C$ & $\varepsilon_\Gamma^D$ & $e \Delta V$\\
    \midrule
    \supercell{4} & 6.294 & 6.246 & 0.024 \\
    \supercell{5} & 6.269 & 6.242 & 0.016 \\
    \bottomrule
    \end{tabular}
    \caption{The intermediate results calculated by us using the input file of \cite{Ma2022, Ma2022_data}.}
    \label{tab:int_res_structure_MA}
\end{table}

\begin{table}[H]
    \centering
    \begin{tabular}{lccc}
    \toprule
    Supercell & $\varepsilon_\Gamma^C$ & $\varepsilon_\Gamma^D$ & $e \Delta V$\\
    \midrule
    \supercell{4} & 6.294 & 6.246 & 0.044 \\
    \supercell{5} & 6.269 & 6.242 & 0.037 \\
    \bottomrule
    \end{tabular}
    \caption{The intermediate results as published in the appendix of \cite{Ma2022}.}
    \label{tab:int_res_MA}
\end{table}

\section{Intermediate results binding energies} \label{apx:intermediate_results}

\begin{table}[H]
    \centering
    \begin{tabular}{llcccc}
    \toprule
    Method & Supercell & $\varepsilon_\Gamma^C$ & $\varepsilon_\Gamma^D$ & $e \Delta V$ & $E_b$ \\
    \midrule
    \multirow{4}{*}{Si P} & \supercell{4} & 6.2801 & 6.2357 & 0.0145 & 0.0589 \\
     & \supercell{5} & 6.2466 & 6.2219 & 0.0165 & 0.0412 \\
     & \supercell{6} & 6.2389 & 6.2212 & 0.0145 & 0.0322 \\
     & \supercell{7} & 6.2418 & 6.2272 & 0.0125 & 0.0271 \\ \midrule
    \multirow{5}{*}{Si-1/4 P-1/2} & \supercell{4} & 6.3895 & 6.3195 & 0.0155 & 0.0855 \\
     & \supercell{5} & 6.3671 & 6.3218 & 0.0165 & 0.0618 \\
     & \supercell{6} & 6.3659 & 6.3294 & 0.0145 & 0.0510 \\
     & \supercell{7} & 6.3733 & 6.3394 & 0.0115 & 0.0454 \\
     & \supercell{8} & 6.3736 & 6.3379 & 0.0085 & 0.0442 \\
    \bottomrule
    \end{tabular}
    \caption{The intermediate results for the calculation of the binding energy in eV for phosphorus.}
    \label{tab:P_int_result}
\end{table}

\begin{table}[H]
    \centering
    \begin{tabular}{llcccc}
    \toprule
    Method & Supercell & $\varepsilon_\Gamma^C$ & $\varepsilon_\Gamma^D$ & $e \Delta V$ & $E_b$ \\
    \midrule
    \multirow{5}{*}{Si As} & \supercell{2} & 6.3876 & 6.0953 & -0.0045 & 0.2878 \\
     & \supercell{3} & 6.3876 & 6.2307 & 0.0175 & 0.1744 \\
     & \supercell{4} & 6.2843 & 6.2169 & 0.0165 & 0.0839 \\
     & \supercell{5} & 6.2508 & 6.2116 & 0.0145 & 0.0537 \\
     & \supercell{6} & 6.2431 & 6.2146 & 0.0135 & 0.0420 \\ \midrule 
    \multirow{5}{*}{Si-1/4 As} & \supercell{2} & 6.5395 & 6.1939 & 0.0145 & 0.3601 \\
     & \supercell{3} & 6.5824 & 6.4640 & 0.0165 & 0.1349 \\
     & \supercell{4} & 6.3940 & 6.3099 & 0.0205 & 0.1046 \\
     & \supercell{5} & 6.3715 & 6.3158 & 0.0165 & 0.0722 \\
     & \supercell{6} & 6.3704 & 6.3249 & 0.0145 & 0.0600 \\ \midrule
    \multirow{7}{*}{Si-1/4 As-1/2} & \supercell{2} & 6.5395 & 6.1214 & -0.0185 & 0.3996 \\
     & \supercell{3} & 6.5824 & 6.4261 & 0.0135 & 0.1698 \\
     & \supercell{4} & 6.3940 & 6.2852 & 0.0165 & 0.1253 \\
     & \supercell{5} & 6.3715 & 6.2973 & 0.0135 & 0.0877 \\
     & \supercell{6} & 6.3704 & 6.3087 & 0.0115 & 0.0732 \\
     & \supercell{7} & 6.3764 & 6.3193 & 0.0075 & 0.0646 \\
     & \supercell{8} & 6.3736 & 6.3155 & 0.0055 & 0.0636 \\ \midrule
    \multirow{4}{*}{Si-1/4 As\_d-1/2} & \supercell{4} & 6.3832 & 6.2721 & 0.0125 & 0.1236 \\
     & \supercell{5} & 6.3607 & 6.2857 & 0.0115 & 0.0865 \\
     & \supercell{6} & 6.3594 & 6.2976 & 0.0105 & 0.0723 \\
     & \supercell{7} & 6.3653 & 6.3085 & 0.0075 & 0.0643 \\
    \bottomrule
    \end{tabular}
    \caption{The intermediate results for the calculation of the binding energy in eV for arsenic.}
    \label{tab:As_int_result}
\end{table}

\begin{table}[H]
    \centering
    \begin{tabular}{llcccc}
    \toprule
    Method & Supercell & $\varepsilon_\Gamma^C$ & $\varepsilon_\Gamma^D$ & $e \Delta V$ & $E_b$ \\
    \midrule
    \multirow{3}{*}{Si Sb\_d} & \supercell{4} & 6.2738 & 6.2370 & 0.0265 & 0.0633 \\
     & \supercell{5} & 6.2402 & 6.2180 & 0.0195 & 0.0417 \\
     & \supercell{6} & 6.2324 & 6.2157 & 0.0175 & 0.0342 \\ \midrule
    \multirow{4}{*}{Si-1/2 Sb-1/2} & \supercell{4} & 6.3940 & 6.3414 & 0.0355 & 0.0881 \\
     & \supercell{5} & 6.3715 & 6.3346 & 0.0235 & 0.0604 \\
     & \supercell{6} & 6.3704 & 6.3393 & 0.0205 & 0.0516 \\
     & \supercell{7} & 6.3764 & 6.3469 & 0.0145 & 0.0440 \\ \midrule
    \multirow{5}{*}{Si-1/2 Sb\_d-1/2} & \supercell{4} & 6.3832 & 6.3263 & 0.0295 & 0.0864 \\
     & \supercell{5} & 6.3607 & 6.3214 & 0.0205 & 0.0598 \\
     & \supercell{6} & 6.3594 & 6.3269 & 0.0185 & 0.0510 \\
     & \supercell{7} & 6.3653 & 6.3351 & 0.0145 & 0.0447 \\
     & \supercell{8} & 6.3736 & 6.3407 & 0.0105 & 0.0434 \\
    \bottomrule
    \end{tabular}
    \caption{The intermediate results for the calculation of the binding energy in eV for antimony.}
    \label{tab:Sb_int_result}
\end{table}

\begin{table}[H]
    \centering
    \begin{tabular}{llcccc}
    \toprule
    Method & Supercell & $\varepsilon_\Gamma^C$ & $\varepsilon_\Gamma^D$ & $e \Delta V$ & $E_b$ \\
    \midrule
    \multirow{5}{*}{Si Bi} & \supercell{2} & 6.3876 & 6.0951 & 0.0745 & 0.3670 \\
     & \supercell{3} & 6.3876 & 6.3554 & 0.0275 & 0.0597 \\
     & \supercell{4} & 6.2843 & 6.2042 & 0.0245 & 0.1046 \\
     & \supercell{5} & 6.2508 & 6.2019 & 0.0185 & 0.0674 \\
     & \supercell{6} & 6.2431 & 6.2047 & 0.0165 & 0.0549 \\ \midrule 
    \multirow{6}{*}{Si-1/4 Bi} & \supercell{2} & 6.5395 & 6.1932 & 0.0665 & 0.4128 \\
     & \supercell{3} & 6.5824 & 6.4417 & 0.0295 & 0.1702 \\
     & \supercell{4} & 6.3940 & 6.2917 & 0.0275 & 0.1298 \\
     & \supercell{5} & 6.3715 & 6.2998 & 0.0175 & 0.0892 \\
     & \supercell{6} & 6.3704 & 6.3083 & 0.0155 & 0.0776 \\
     & \supercell{7} & 6.3764 & 6.3176 & 0.0105 & 0.0693 \\ \midrule
    \multirow{6}{*}{Si-1/4 Bi-1/2} & \supercell{2} & 6.5395 & 6.1153 & 0.0705 & 0.4947 \\
     & \supercell{3} & 6.5824 & 6.3878 & 0.0275 & 0.2221 \\
     & \supercell{4} & 6.3940 & 6.2528 & 0.0225 & 0.1637 \\
     & \supercell{5} & 6.3715 & 6.2688 & 0.0145 & 0.1172 \\
     & \supercell{6} & 6.3704 & 6.2799 & 0.0125 & 0.1030 \\
     & \supercell{7} & 6.3764 & 6.2898 & 0.0075 & 0.0941 \\ \midrule
      \multirow{5}{*}{Si-1/4 Bi\_d-1/2} & \supercell{4} &                  6.3940 &                  6.2677 &        0.0285 & 0.1548 \\
    & \supercell{5} & 6.3715 & 6.2781 & 0.0175 & 0.1109 \\
    & \supercell{6} & 6.3704 & 6.2879 & 0.0135 & 0.0960 \\
    & \supercell{7} & 6.3764 & 6.2974 & 0.0095 & 0.0885 \\
    & \supercell{8} & 6.3857 & 6.3041 & 0.0065 & 0.0881 \\ \midrule
\multirow{3}{*}{Si-1/4 Bi\_d-1/2 (SOC)} & \supercell{4} &                  6.3940 &                  6.2707 &        0.0265 & 0.1498 \\
    & \supercell{5} & 6.3715 & 6.2805 & 0.0175 & 0.1085 \\
    & \supercell{6} & 6.3704 & 6.2902 & 0.0105 & 0.0907 \\
    \bottomrule
    \end{tabular}
    \caption{The intermediate results for the calculation of the binding energy in eV for bismuth.}
    \label{tab:Bi_int_result}
\end{table}

\end{appendices}
\end{document}